\documentclass[reprint,aps,prb,amsmath,amssymb,twocolumn,showkeys, superscriptaddress,floatfix,showkeys,twoside]{revtex4-2}
\usepackage{graphicx}
\usepackage{dcolumn}
\usepackage{bm}
\usepackage[colorlinks = true, allcolors = blue]{hyperref}
\usepackage{verbatim}
\usepackage{soul}
\usepackage{float}
\usepackage{xcolor}
\usepackage{multirow}
\usepackage{fancyhdr}
\begin{document}

\title{Thermal effect on magnetoexciton energy spectra in monolayer \\transition-metal dichalcogenides}

\author{Duy-Nhat Ly}
\email{nhatld@hcmue.edu.vn}
\affiliation{Computational Physics Key Laboratory K002, Department of Physics, Ho Chi Minh City University of Education, Ho Chi Minh City 72759, Vietnam}
\thanks{These authors contributed equally to this paper.}

\author{Dai-Nam Le}
\email{dainamle@usf.edu}
\affiliation{Department of Physics, University of South Florida, Tampa, FL 33620, United States of America}
\thanks{These authors contributed equally to this paper.}

\author{Ngoc-Hung Phan}
\affiliation{Computational Physics Key Laboratory K002, Department of Physics, Ho Chi Minh City University of Education, Ho Chi Minh City 72759, Vietnam}

\author{Van-Hoang Le}
\email{hoanglv@hcmue.edu.vn}
\affiliation{Computational Physics Key Laboratory K002, Department of Physics, Ho Chi Minh City University of Education, Ho Chi Minh City 72759, Vietnam}

\date{April 07, 2023}

\begin{abstract}

It is widely comprehended that temperature may cause phonon-exciton scattering, enhancing the energy level's linewidth and leading to some spectrum shifts. However, in the present paper, we suggest a different mechanism that allows the thermal motion of the exciton's center of mass (c.m.) to affect the magnetoexciton energies in monolayer dichalcogenides (TMDCs). By the nontrivial but precise separation of the c.m. motion from an exciton in a monolayer TMDC with a magnetic field, we obtain an equation for the relative motion containing a motional Stark term proportional to the c.m. pseudomomentum, related to the temperature of the exciton gas but neglected in the previous studies. Solving the Schr{\"o}dinger equation without omitting the motional Stark potential at room temperature shows approximately a few meV thermal-magnetic shifts in the exciton energies, significant enough for experimental detection. Moreover, this thermal effect causes a change in exciton radius and diamagnetic coefficient and enhances the exciton lifetime as a consequence. Surprisingly, the thermoinduced motional Stark potential breaks the system's SO(2) symmetry, conducting new peaks in the exciton absorption spectra at room temperature besides those of the $s$ states. This mechanism could be extended for other magnetoquasiparticles such as trions and biexcitons.\\

Preprint of \href{https://journals.aps.org/prb/abstract/10.1103/PhysRevB.107.155410}{Phys. Rev. B \textbf{107}, 155410 (2023)} \cite{Ly2023a}.
\end{abstract}


\maketitle

\section{Introduction} 
During the last two decades, monolayer transition-metal dichalcogenides (TMDCs) have become hot spots for papering the formation of excitons because of their unique property of electron-hole interaction \cite{berkelbach2013, chernikov2014,Wang2018, Molas2019}. Unlike three-dimensional excitons in novel semiconductors, neutral excitons in these monolayer TMDCs are thermally stable at room temperature, even for the Rydberg states. Their high binding energy provides beneficial optical properties in both ultraviolet (UV) and infrared (IR) ranges \cite{He2014, Ugeda2014, Ye2014, Hill2015, Arora2015}. Especially, exciton energy spectra under the presence of a constant magnetic field have been of great interest recently both in experimental observations and theoretical studies due to their incredible potential in retrieving several physical quantities of monolayer TMDCs, such as effective masses of electrons and holes, the Land\'e $g$ factors, and two-dimensional polarizability \cite{Aroga2019,Stier2016-nat, Stier2016-nano, Plechinger2016, VanDerDonck2018, Stier2018, Zipfel2018, Chen2019-nano, Nguyen2019, Liu2019, Goryca2019, Kezerashvili2021, ly2023retrieval}. Therefore, it is essential to qualitatively (and quantitatively, if possible) comprehend the influences of external factors like temperature on the energy spectra of magnetoexcitons in the monolayer TMDCs.

\begin{figure}[hb]
\begin{center}
\includegraphics[width=0.75 \columnwidth]{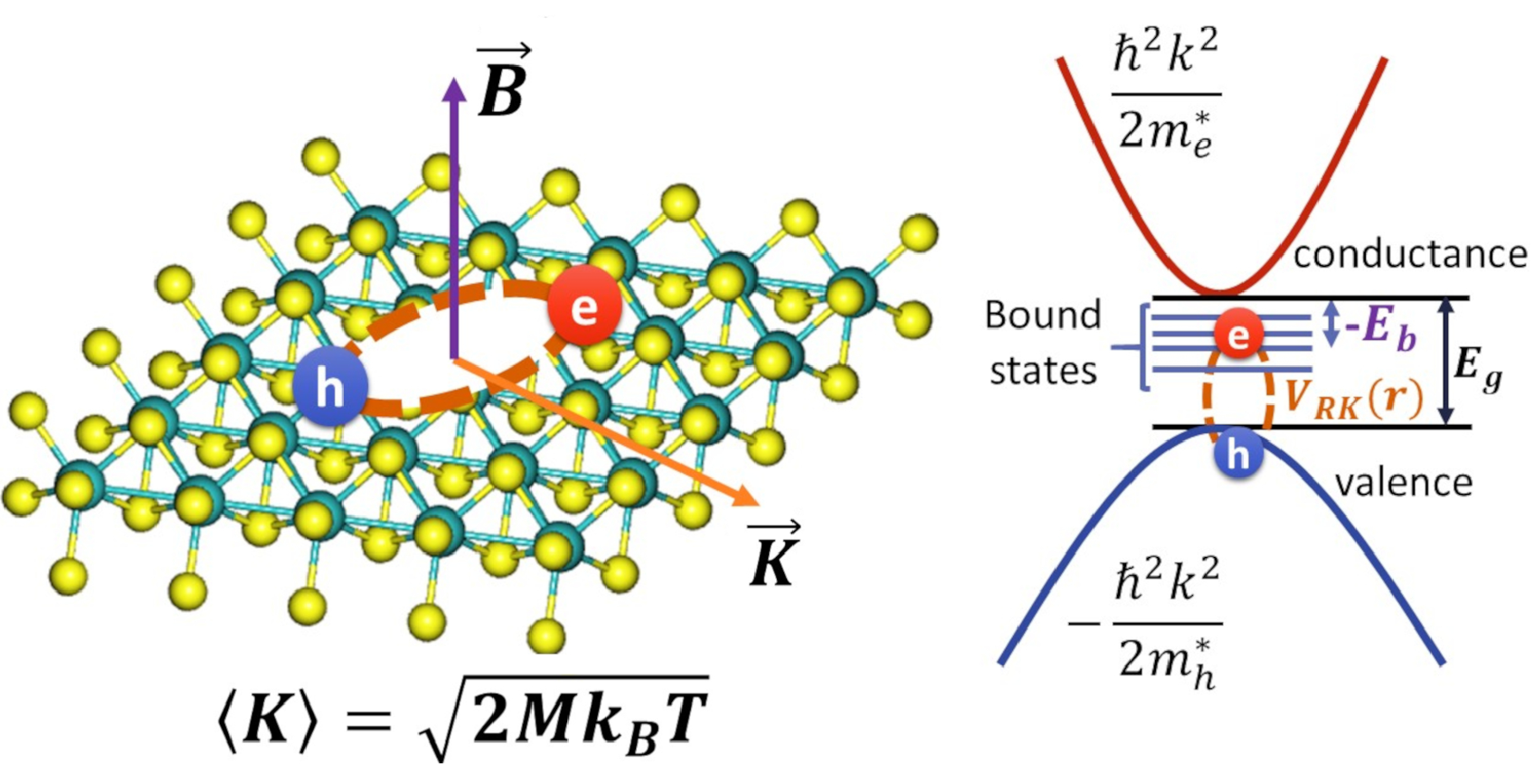}
\caption{(Color online) Excitons in a monolayer TMDC with a magnetic field. The c.m. pseudomomentum is conserved and related to the thermal motion.}
\label{fig1}
\end{center}
\end{figure}

The thermal effect on the exciton energies has been intensively investigated recently and explained by the exciton-phonon scattering \cite{Lengers2021, Henriques2021}. Indeed, the quantization of the crystal lattice oscillation results in quasiparticles named \textit{phonons} which carry the thermal motion. The scattering between excitons and phonons affects the relative motion of the electron-hole pair that, consequently, causes some energy shifts of the exciton bound states. Recently, experiments for exciton energies in monolayer TMDCs have been conducted with a high magnetic field \cite{Liu2019, Goryca2019, Kezerashvili2021}, raising another question related to the center of mass (c.m.) separation. It is well known that the c.m. of a two-body system in a magnetic field can be separated, but the c.m. pseudomomentum remains in the equation for relative motion \cite{comment2020, Avron1978, Ruder1994}. This circumstance leads to a mechanism that the thermal motion of the c.m. can affect the exciton energies. We will show in the present paper additional energy shifts caused by this thermal-magnetic effect.

The exciton-phonon scattering at finite temperatures affects the imaginary part of exciton energy that makes exciton decay \cite{Toyozawa2003, Chermikov2012, Palummo2015, Selig2016, Wang2018, Dinh2018, Brem2019, Roux2021}. However, in our thermal-magnetic mechanism, the motional Stark potential, linearly proportional to the magnetic intensity, could be more inconspicuous than the diamagnetic potential depending on the squared magnetic intensity and may contribute only to the energy shift but not to the exciton lifetime. Nevertheless, this thermoinduced potential could make the average radius of the electron-hole pair larger and consequently enhance the radiative lifetime of excitons in monolayer TMDCs \cite{Chermikov2012, Brem2019, Roux2021}. Besides, the symmetry breaking of the system due to the additional thermoinduced term can affect the wave functions. Therefore, we aim to examine the effect of finite temperature on the exciton energy spectra, the diamagnetic coefficient, and the exciton radius as a function of magnetic intensity and their consequence on the absorption spectra and the exciton radiative lifetime.

The rest of this paper is as follows. Section \ref{Sec2} presents the theoretical background for our paper, including the separation of the c.m. motion, the introduction of the motional Stark potential, and the assumption about its temperature dependence. Then, the results and discussion are given in Sec. \ref{Sec3} with the thermal effect on the diamagnetic coefficients, the thermal-magnetic shift on exciton energies, the thermoinduced symmetry-breaking and enhancement of exciton lifetime, and the new peaks in absorption spectra. Section \ref{Conc} includes our conclusion.

\section{Theoretical background}\label{Sec2}
\subsection{Schr{\"o}dinger equation} 
To consider the thermal effect, we need to solve the Schr{\"o}dinger equation of the electron-hole pair in the framepaper of the effective mass approximation, where a Wannier exciton in a monolayer TMDC is described as a two-dimensional system of one electron and one hole interacting with each other by the potential ${\hat V}_{h-e}(r)$ in the $xy$ plane, as shown in Fig.~\ref{fig1}. Because of the two-dimensional many-particle effect, the interaction potential ${\hat V}_{h-e}$ is no longer Coulombic but screened and described by the Rytova-Keldysh potential \cite{Rytova1967,keldysh1979,berkelbach2013,chernikov2014, cudazzo2011, haramura1988}. One required task before solving the Schr{\"o}dinger equation is to separate the movement of the electron-hole c.m with the coordinate $\mathbf{R}$. However, for a two-body system such as an exciton in a magnetic field $\mathbf{B} = B \mathbf{e}_z$, the c.m. dynamics can not be removed totally from the equation describing the electron-hole relative motion. The variable-separation procedure for this system is no longer trivial because the magnetic field breaks the system's translation symmetry, which leads to the nonconservation of the total momentum $\mathbf{P}$ of the electron-hole pair. Fortunately, there is instead another constant of motion, pseudomomentum $\hat{\mathbf{P}}_{0} = \hat{\mathbf{P}}- \frac{1}{2} e \mathbf{B}\times \mathbf{r}$, commuting with the system's Hamiltonian, i.e., $[\hat{\mathbf{P}}_0, {\hat H}_{X}]= 0$. Using this conservative quantity, one can obtain an equation for the relative motion \cite{comment2020, Avron1978, Ruder1994}. In this case, the wave function can be written as
\begin{equation}\label{eq5}
\Psi_{\mathbf{K}} (\mathbf{R},\mathbf{r}) = {e^{\frac{i}{\hbar }\left( {\mathbf{K} + \frac{1}{2}e\mathbf{B} \times \mathbf{r}} \right) \cdot \mathbf{r}}}\,\psi_{\mathbf{K}} (\mathbf{r}),
\end{equation}
where $\mathbf{K}$ is an eigenvector of operator $\hat{\mathbf{P}}_0$. The wave function for relative motion $\psi_{\mathbf{K}} (\mathbf{r})$ is obtained by solving the Schr{\"o}dinger equation
\begin{eqnarray}\label{eq6}
\left\{ \frac{\hat{\mathbf{p}}^2}{2 \mu} + \frac{1-\sigma}{1+\sigma} \right. \frac{{eB}}{2\mu}{{\hat l}_z}
 + \frac{{e^2} {B^2}}{8 \mu } {r}^2 + {\hat V}_{h-e}(r) \quad\quad\quad\nonumber\\
\left. - \frac{e}{M}({\mathbf{B}\times\mathbf{K}})\cdot \mathbf{r} - E \right\} \psi_{\mathbf{K}} (\mathbf{r}) = 0.
\end{eqnarray}
with the total mass $M=m^{*}_e+m^{*}_h$, the exciton reduced mass $\mu=m^{*}_e m^{*}_h/(m^{*}_e+m^{*}_h)$, and the ratio of masses $\sigma=m^{*}_e/m^{*}_h$. Here, $m^{*}_e$ and $m^{*}_h$ are the effective masses of the electron and hole; $e$ is the elementary charge with the positive value; $\hat{\mathbf{p}}$ and $\hat l_z$ are the operators of the momentum and angular momentum of the relative motion. A detailed derivation of Eq.~\eqref{eq6} is given in Suppl.~\cite{Suppl}.

For the exciton in monolayer TMDCs such as $\textrm{WSe}_2$ as considered in this paper, the exciton reduced mass $\mu=0.20 \times m_e$, mass ratio $\sigma=0.94$, total mass $M=0.80 \times m_e$  ($m_e$ is electron mass), and the Rytova-Keldysh potential's parameters (the average dielectric constant of the surrounding material $\kappa=4.5$, screening length $r_0=4.21$) are taken from Refs. \cite{Stier2018, Goryca2019}.

\subsection{Thermoinduced motional Stark potential} 
In equation~\eqref{eq6}, we consider the specific term $ V_{mS}=- \frac{e}{M}({\mathbf{B}\times\mathbf{K}})\cdot \mathbf{r}$ containing the c.m. pseudomomentum $\mathbf{K}$  and will show its relation to the temperature of the neutral exciton gas. First, we note that the density of excitons $n_X$ in monolayer TMDCs is low, i.e., the average distance between two excitons $1/\sqrt{n_X}$ is much larger than the thermal wavelength $\lambda_{th} = \sqrt{\frac{2\pi\hbar^2}{M k_B T}}$ so that the Maxwell-Boltzmann statistics is still valid. Indeed, the density $n_X$, i.e., the number of neutral excitons per unit of area, could be modulated around $10^{11} \text{--} 10^{12} \text{ cm}^{-2}$ in real experiments for monolayer WSe$_2$ \cite{Poellmann2015, Yan2015, You2015, Steinleitner2017}, which is much smaller than its limit at room temperature $\lambda_{th}^{-2}= 4.3 \times 10^{12} \text{ cm}^{-2}$. Then, we estimate the root-mean-square of the c.m. pseudomomentum at temperature $T$ by the equipartition theorem as 
$\frac{1}{2M}  \overline{\text{K}^2} =  k_B T$, where $k_B$ is the Boltzmann constant. Instead of calculating exciton energies at each pseudomomentum value and then averaging them as $\overline{E(K)}$, we do it another way by calculating the exciton energy $ E(\overline{\text{K}}) $ at the average pseudomomentum value. The validity of this approach has been confirmed numerically in Tab.\ref{tabS1a} in Suppl.~\cite{Suppl}. This approach allows the considered potential to relate to the temperature as follows
\begin{eqnarray}\label{eq6c}
V_{mS} (\mathbf{r}) =- \sqrt{\frac{2k_B T}{M}} eB\, x .
\end{eqnarray}
Here, without losing generality, we consider the case where $\mathbf{K}$ is along the $y$ axis so that vector $\mathbf{B} \times \mathbf{K}$ is along the $x$ axis. A more microscopic grounding for the thermoinduced term \eqref{eq6c} is given in Sec.\ref{sec:S1new} and Sec.\ref{sec:S2new} in Suppl.~\cite{Suppl}.  

The potential \eqref{eq6c} influences the exciton in the same way as the Stark effect \cite{Farrelly1994, Schweiner2017}, so we call it \textit{\textcolor{blue}{thermoinduced motional Stark potential}}. Recent theoretical studies of magnetoexcitons in monolayer TMDCs  always neglect this potential  \cite{VanDerDonck2018, Nguyen2019, Kezerashvili2021} that is reasonable only at zero temperature. However, the experimental observations were conducted not only at low temperatures \cite{Stier2016-nat, Plechinger2016, Zipfel2018, Stier2018, Liu2019, Goryca2019, Chen2019-nano} but also at room temperature \cite{Arora2015, Aroga2019,Lengers2021, Henriques2021}. Hence, the motional Stark effect arising from the thermal fluctuation of the c.m. pseudomomentum needs to be considered when calculating the energy spectra of magnetoexcitons in monolayer TMDCs.

We can see that the diamagnetic term $V_{diamag.} (\mathbf{r}) =\frac{{e^2} {B^2}}{8 \mu } {r}^2$ in the effective potential of Eq.~\eqref{eq6} is quadratically proportional to the electron-hole distance and consequently dominant at the large separation of the electro-hole pair compared to the thermoinduced Stark term, which is linearly proportional to the variable $x$. As a result, there is no tunneling effect, even considering the thermoinduced motional Stark potential. The exciton is always in its bound states, unlike the well-known LoSurdo-Stark effect in the two-dimensional electron gas \cite{Tanaka1987,Henriques2020} and field-induced dissociation of excitons in a TMDC \cite{Kamban2019}. Instead of tunneling, we expect the thermoinduced motional Stark potential could cause the Stark shift in the energy spectra.

\section{Results and Discussion}\label{Sec3}
\subsection{Thermal effect on diamagnetic coefficient and Landau levels} 
To see how the thermoinduced motional Stark potential affects the asymptotic behaviors of exciton energies in the limits of low and high magnetic intensities, we calculate the energies analytically by applying the perturbation theory to the Schr{\"o}dinger equation \eqref{eq6}.
For low magnetic intensity, restricted by the condition that the typical length in the magnetic field  is much larger than the average exciton radius: $l_B = \sqrt{{\hbar}/{e B}}\gg \left< r \right>_{nm}$, we have the energy of the ($n,m$) quantum state in the second perturbation order as
\begin{eqnarray}\label{eqn9a}
E^{(2)}_{nm} (B, T)  = E_{nm}^{(0)}  + \frac{1 - \sigma}{1 + \sigma} \frac{m \hbar}{2 \mu}\, eB \quad\quad\nonumber\\
 + \frac{\left< r^2 \right>_{nm}} {8 \mu}\, e^2B^2 - \alpha_{nm}\frac{k_B T }{M} \,e^2 B^2,
\end{eqnarray}
where $E^{(0)}_{nm}$, $\left< r^2 \right>_{nm}$, and $\alpha_{nm}$ are the zero-field energy, squared radius, and polarizability of the exciton. 
To calculate these quantities, we need to solve the Schr{\"o}dinger equation in the zeroth order of approximation. However, the solutions cannot be obtained analytically, so, we do it another way by numerically solving equation \eqref{eq6} and then fitting the obtained results to formula \eqref{eqn9a}. The concrete values are presented in Table \ref{tab1a} for $1s$, $2s$, $3s$, $2p_{-1}$, and $2p_{+1}$ states. 

\begin{table}[htbp]
	\caption{\label{tab1a} Zero-field energies, squared radius, and polarizability.}
	\begin{ruledtabular}
		\begin{tabular}{l r r r r r r}
				&	$1s$	&	$2s$	& $3s$&	$2p_{-1}$	&	$2p_{+1}$	\\
			\hline
			$E^{(0)}_{nm}$  (eV)	&	-0.1686	&	-0.0386	&	-0.0166	&	-0.0498	&	-0.0498	\\
$\left< r^2 \right>_{nm}$ (nm$^2$)	&	2.63	&	45.8	&	241	&	21.1	&	21.1	\\
$\alpha_{nm}$ (nm$^2$/eV)	&	3.91	&	-112	&	-6060   	&	259	&	240	\\
		\end{tabular}
	\end{ruledtabular}
\end{table}

The last term in the exciton energy \eqref{eqn9a}, the motional Stark correction, is quadratically proportional to the magnetic field (see also in Fig.\ref{figS2} in Suppl. \cite{Suppl}). That means the thermal effect contributes to the diamagnetic coefficient $\sigma_{nm}$ defined by the equation 
\begin{eqnarray}
\label{eqn9e}
\sigma_{nm} (T) && = \lim\limits_{B \to 0}\frac{1}{2} \frac{\partial^2 E_{nm}(B,T)}{\partial B^2} \nonumber\\
&& = \frac{e^2}{8 \mu} \left< r^2 \right> _{nm} - \frac{ e^2}{M} \alpha_{nm} k_B T.
\end{eqnarray}
Table \ref{tab2a} shows some values for the zero-temperature diamagnetic coefficient  $\sigma^{0}_{nm} =\frac{e^2}{8 \mu} \left< r^2 \right> _{nm}$ and their thermoinduced corrections $\Delta \sigma_{nm}=- \frac{ e^2}{M} \alpha_{nm} k_B T$ for the $1s$, $2s$, $3s$, $2p_{-1}$, and $2p_{+1}$ states at room temperature. The thermoinduced corrections are about $8\%$ for $1s$, $13\%$ for $2s$, and extremely high for higher states: $2p_{+1}$, $2p_{-1}$, and $3s$  ($56\%$, $60\%$, and $130\%$). These are significant enough to impact experimental measurement that we should pay attention to them when measuring exciton energies at finite temperatures.

\begin{table}[htbp]
	\caption{\label{tab2a} Zero-temperature diamagnetic coefficients and their thermoinduced corrections at room temperature in unit of $\mu$eV/Tesla$^2$ compared to the experimental data \cite{Stier2018}.}
	\begin{ruledtabular}
		\begin{tabular}{r r r r r r r}
				&	$1s$	&	$2s$	& $3s$&	$2p_{-1}$	&	$2p_{+1}$	\\
			\hline
     $\sigma_{nm}$	\cite{Stier2018} &	$0.31\pm0.02$ 	&	$4.6\pm 0.2$	&	$22\pm 2$	&	-	&	-	\\
   $\sigma^{0}_{nm}$	&	0.289	&	5.039	&	26.53	&	2.429	&	2.429	\\
$\Delta \sigma_{nm}$	&	-0.022	&	0.637	&	34.42	&	-1.469	&	-1.361	\\
$\frac{|\Delta \sigma_{nm}|}{\sigma^{0}_{nm}} $	&	8\%	&	13\%	&	130\%	&	60\%	&	56\%	\\

		\end{tabular}
	\end{ruledtabular}
\end{table}

For high magnetic intensity, restricted by condition  $l_B =\sqrt{\hbar/eB} \ll \left< r \right>_{nm}$, the harmonic oscillator potential becomes dominant compared to the Rytova-Keldysk potential. In this case, the main part of the Hamiltonian is the magnetic term; therefore, we can obtain the Landau levels for energies with the thermoinduced Stark corrections by the perturbation theory as
\begin{eqnarray}
\label{eqn9c}
E^{(2)}_{nm} (B, T) = && \frac{\hbar }{2 \mu} \left( 2 n - |m| - 1 + \frac{1-\sigma}{1+\sigma} m  \right) e B \nonumber\\
&& \quad\quad - \frac{8 \mu \beta_{nm}}{M} k_B T ,
\end{eqnarray}
where the dimensionless coefficients $\beta_{nm}$ are independent of the temperature $T$, see details in Suppl.~\cite{Suppl}. Because of the neglect of Coulomb interaction, the energy formula \eqref{eqn9c} is valid only for very high magnetic intensity, extremely higher than the laboratory limit of about 91 Tesla \cite{Goryca2019}. Nevertheless, this formula gives the right rule for thermoinduced shifts from the Landau levels $\Delta E= - \frac{8\mu\beta_{nm}}{M} k_B T$, i.e., linearly proportional to the temperature and almost independent of the magnetic field (see also in Fig.\ref{figS3} in Suppl. \cite{Suppl}). 

\subsection{Thermal-magnetic shift on exciton energies} 
\begin{figure}[htbp]
\begin{center}
\includegraphics[width=1.0 \columnwidth]{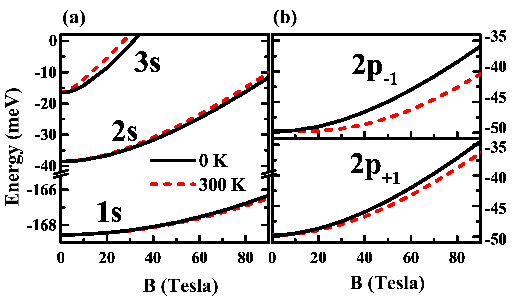}
\caption{(Color online) Exciton energies for the $1s$, $2s$, $3s$ (a), and $2p_{-1}$, $2p_{+1}$ (b) states depending on the magnetic intensity $B$ at temperatures 0~K and 300~K.}
\label{fig2}
\end{center}
\end{figure}

We now numerically investigate the thermal effect on energy by the Feranchuk-Komarov operator method \cite{Hoangbook2015, Nguyen2019}. The obtained energies are given in Tab.\ref{tabS1} -- Tab.\ref{tabS5} in Suppl. \cite{Suppl} for the $1s$, $2s$, $3s$, $2p_{-1}$, and $2p_{+1}$ states. For an illustration of the effect, Figs.~\ref{fig2} (a) and (b) show exciton energies depending on the magnetic intensity at 0~K and 300~K. 

We see that the temperature does not influence the $1s$ state much. At the same time, the thermal effect is noticeable for the $2s$ and higher Rydberg states for the magnetic intensity of more than 60 Tesla. Particularly for the magnetic intensity of 90 Tesla, the energy shifts when including the temperature of 300 K are shown in Table~\ref{tab3a}. These shifts are big enough compared to the experimental sensitivity of 1 meV and are caused by the thermal motion of excitons in a magnetic field; thus, we call them the \textit{\textcolor{blue}{thermal-magnetic shifts}}.

\begin{table}[htbp]
	\caption{\label{tab3a} Thermal-magnetic shifts in exciton energy spectra calculated between the temperatures of 0 K and 300 K for the magnetic field of 90 Tesla.}
	\begin{ruledtabular}
		\begin{tabular}{l r r r r r r}
				&	$1s$	&	$2s$	& $3s$&	$2p_{-1}$	&	$2p_{+1}$	\\
			\hline
			$\Delta E $  (meV)	&	0.2	&	1.5	&	3.9	&	4.5	&	1.9	\\
		\end{tabular}
	\end{ruledtabular}
\end{table}

We note that the thermal-magnetic shift predicted above is comparable with the shift caused by the phonon-exciton interaction \cite{Lengers2021, Henriques2021}. For example, the polaron shift is about 15 meV at room temperature for $1s$ state exciton in the monolayer TMDC as shown in Ref.~\cite{Henriques2021}. Compared with this, the thermal-magnetic shift for the 1s state, which is 0.2 meV as shown in Table \ref{tab3a}, can be ignored. It means that neglecting the exciton c.m. motion in Ref. \cite{Henriques2021} is feasible in this case. However, for a higher excited state such as $3s$, the thermal-magnetic shift of 3.9 meV should be taken into account if the polaron shift are calculated within the presence of a high magnetic field. The two shifts are from different mechanisms (phonon-exciton scattering versus the thermal motion of the exciton c.m. in a high magnetic field). They can be calculated separately, and both deserve consideration in analyzing experimental data for excited states in a high magnetic field. Also, for a high magnetic field of about 30 Tesla, as considered in Ref.~\cite{Robert2020}, the fine structure energy split of excitons in monolayer TMDCs caused by the spin-magnetic interaction is up to tens of meV. Compared with this, the thermoinduced shift in the present paper is relatively noticeable, and it is important to consider.

\subsection{Thermoinduced symmetry-breaking and enhancement of exciton lifetime} 
\begin{figure}[htbp]
\begin{center}
\includegraphics[width=0.95 \columnwidth]{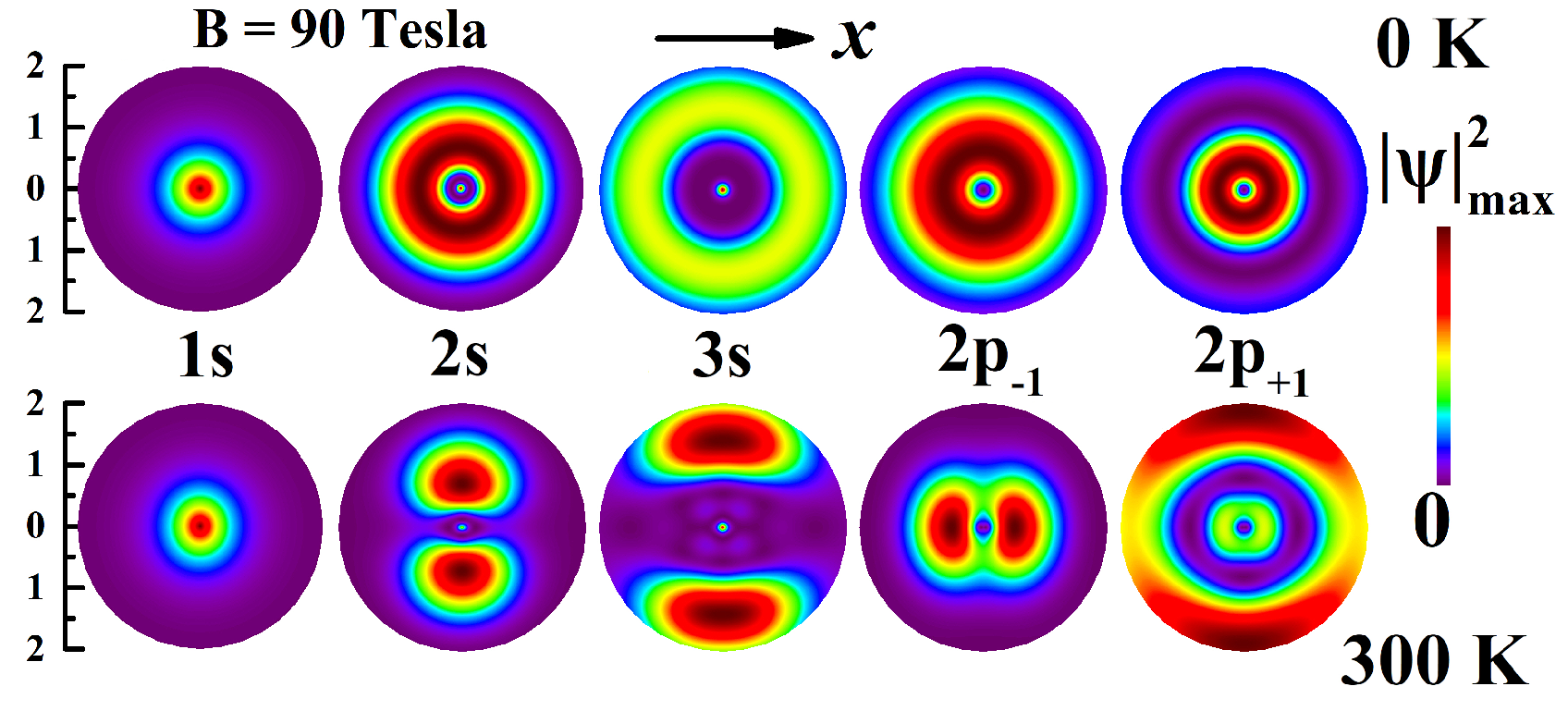}
\caption{(Color online) Deformation of wave functions due to the thermoinduced symmetry breaking.}
\label{fig3}
\end{center}
\end{figure}

It is well-known that the system of a two-dimensional exciton in a magnetic field, perpendicular to the monolayer TMDC plane, has the potential energy dependent on radius $r=\sqrt{x^2+y^2}$ only and consequently possesses the SO(2) symmetry.  However, if included, the thermo-reduced motional Stark potential \eqref{eq6c}, which depends on the angle $\varphi$ as $\sim r \cos{\varphi}$, will break this symmetry. In this case, the angular momentum $l_z$ is not conserved, and the magnetic number $m$ is no longer a good quantum number. The consequence is that there are no true $s$ states anymore, but only mixed states with $m\neq 0$ from the basis set functions; see Figs.~\ref{fig3} for the thermoinduced deformation of wave functions. More about the symmetry-breaking effect on the wave function deformation can be found in Sec.\ref{sec:S2C} in Suppl.~\cite{Suppl}. 

The wave function deformation leads to the change of the exciton radius, essential for the exciton radiative lifetime $\tau_{rad}$, which is related to the average electron-hole distance $\left< r\right>$ by the scaling law  $\tau _{rad} \sim \left< r \right>^{\xi}$ \cite{Toyozawa2003,Roux2021}. The scaling factor $\xi$ mainly depends on the dimensionality of the exciton. We roughly take the value of $\xi \approx 3.5$ extracted by studying the exciton in hBN, diamond, and silicon provided in Ref. \cite{Roux2021}.

To have an analytical estimation, we get the formula for the exciton radius of $s$ states by the perturbation theory in the low magnetic field limit as 
\begin{equation}\label{eq22a}
\left< r \right>_{T}=\left< r \right>+g \,e^2B^2\frac{k_B T}{M},
\end{equation}
where the radius $\left< r \right>_T$ is calculated using the wave functions with thermal effect. In contrast,  the free-field wave functions are used to calculate $\left< r \right>$. The coefficient $g$ has the following values $0.17 \times 10^{-3}$, $0.04$, and $-7.5$ (nm$^3$/meV$^2$) for $1s$, $2s$, and $3s$ states, respectively. Consequently, we can get the following formula 
\begin{equation}\label{eq23a}
\frac{\Delta \tau _{rad}}{\tau _{rad}}=\xi \,\frac{\Delta\left< r \right> }{\left< r \right>}
\end{equation}
for the thermal correction $\Delta \tau _{rad}$ to the exciton radiative lifetime, where $\Delta\left< r \right>=g e^2B^2\frac{k_B T}{M}$ for the low magnetic intensity. The correction $\Delta\left< r \right>$ should be calculated numerically for the high magnetic field. We estimate the ratio \eqref{eq23a} for the $1s$, $2s$, and $3s$ states at room temperature and magnetic field of $90$ Tesla and get the following values: $1.4\%$, $4.9\%$, and $2.1\%$, relatively considerable. The thermoinduced correction to the radiative lifetime \eqref{eq23a} qualitatively agrees with the first-principle calculations and experimental observations given in Refs. \cite{Palummo2015, Selig2016, Brem2019} for the $1s$ state, meaning that it always enhances the lifetime and is linearly proportional to the temperature. The thermal motion also causes the lifetime to be enhanced for the $2s$ state but reduced for the $3s$ state. Therefore, this effect (enhancement/reduction of the lifetime) needs further investigation for higher states in our next paper.

\subsection{New peaks in absorption spectra due to the symmetry-breaking} 
\begin{figure}[htbp]
\begin{center}
\includegraphics[width=1.0 \columnwidth]{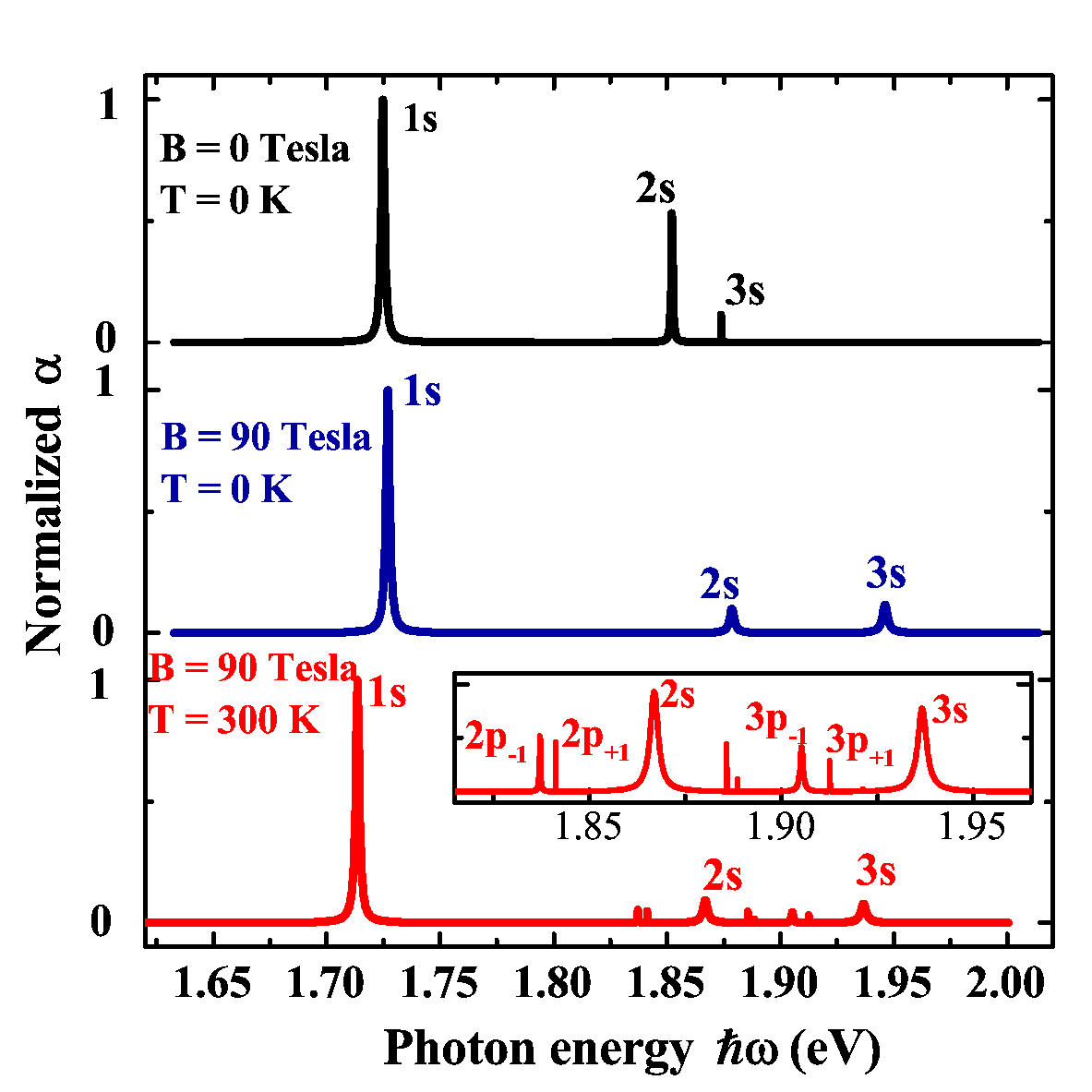}
\caption{(Color online) Normalized optical absorption spectra at temperature of 0 K (black and dark blue) and at room temperature (red).}
\label{fig4}
\end{center}
\end{figure}

Interestingly, the system's symmetry-breaking at room temperature can lead to new peaks in the magnetoexciton absorption spectra, as shown in Fig.~\ref{fig4}. First, we calculate the imaginary part of the susceptibility by the Elliot formula as 
\begin{eqnarray}\label{eqn23}
\alpha (\omega) = C\, \text{Im}\sum_{n,m} \frac{ \omega\,\left| \psi_{nm}(\mathbf{r} = 0)\right|^2}{E_{nm} + E_g - \hbar \omega + i \hbar \tau ^{-1} }
\end{eqnarray}
based on the linear response theory of the exciton \cite{Kira2006, Haug2009, Wu2019, Wu2021}. We use this quantity to estimate the absorption spectra since they are proportional. Here, the coefficient $C$ depends on the materials' background refractive index and interband transition dipole matrix elements. We are only interested in the general picture and thus choose $C$ to normalize the free-field $1s$ state peak to 1 and then use it as a constant for all other states. 

In Eq.~\eqref{eqn23}, the bandgap energy is taken by $E_g=$ 1890 meV from the experiment \cite{Stier2018}. For calculation at room temperature, we subtract a value of 65 meV from the bandgap, contributed by the Varshni and polaron shifts, based on the recent paper \cite{Henriques2021}. Besides, we also add the c.m. kinetic energy $K^2/2M$ of 51.8 meV to the relative energy to get the total exciton energy $E_{nm}$. Regarding the total lifetime, one usually considers both radiative and non-radiative dephasing effects as ${1}/{\tau} = {1}/{\tau_{rad}} + {1}/{\tau_{non.rad}}$ . However, we roughly estimate the exciton lifetime $\tau$ by the scaling law $\tau _{rad} \sim \left< r \right>^{\xi}$ \cite{Toyozawa2003,Roux2021} and fit the coefficient from the experiment data for the $1s$ state \cite{Palummo2015}, which suggests the exciton lifetime around $1$ ps. The thermal correction to the lifetime based on the formula~\eqref{eq23a} of about a few percent is not noticeable in Fig.~\ref{fig4}.

Figure~\ref{fig4} illustrates the normalized linear optical absorption spectra for the quantum states with principal quantum numbers $n<4$. This qualitative picture shows the appearance of new peaks when considering the thermal effect. The explanation is based on the deformation of the wave functions, noting that the degeneracy lifting of the excited states and shifts of energies by the magnetic field and the thermoinduced potential also play an important role. Indeed, since the oscillation strength related to these peaks is proportional to the squared modulus of the wave functions at zero electron-hole separation, only the states associated with $\left| \psi_{nm}(\mathbf{r} = 0)\right|^2 \neq 0$ can be determined by the linear optical absorption spectrum. For zero-temperature magnetoexciton, only the $s$ state wave functions have non-vanished oscillation strength while they vanish for all the other states such as $p$ and $d$; hence, we can only detect the exciton $s$ states from the linear optical response, see Fig.~\ref{fig4} (black and dark blue lines). However, because of the thermal effect, the $p$ and $d$ states now become the superposition of other states, including $s$ states. Therefore, we get the signal of the $p$- and $d$-state-exciton peaks on the linear optical absorption spectra at room temperature, as shown in Fig.~\ref{fig4} (red line). One can see more details about analytical examination of non-$s$-state peak emergence in Sec.\ref{sec:S2D} in Suppl. \cite{Suppl}.

\section{Conclusion}\label{Conc}
By separating the center of mass motion from an exciton in monolayer transition-metal dichalcogenide  WSe$_2$ with a magnetic field, we have pointed out the thermoinduced motional Stark potential on the Schr{\"o}dinger equation, which was previously neglected. Based on this, we have proposed a new mechanism that the thermal motion of the exciton c.m. in a magnetic field could affect the energy spectra of the magnetoexciton. As an observation from our calculation, the thermal-magnetic shifts in the energy spectra are comparable with the polaron shifts caused by the exciton-phonon interaction, thus, should be considered for magnetoexciton energies at room temperature and can be observed by the shifts of resonance peaks on the absorption spectra. The thermoinduced potential also affects the diamagnetic coefficient and breaks the system's SO(2) symmetry. The symmetry-breaking leads to the orbital deformation of the magnetoexciton that indirectly changes the exciton radiative lifetime, which can be observed by examining the width of the resonance peaks on the absorption spectra. Surprisingly, the system's symmetry-breaking at room temperature also causes the $p$-state exciton peaks to emerge on the linear optical absorption spectra, which cannot happen at zero temperature. These results provide another aspect of studying the influence of temperature on magnetoexciton and could be extended for other monolayer TMDCs such as $\text{WS}_2$.  As an outlook, this mechanism could be applied to trions, biexcitons in Dirac materials, and magnetoexcitons originating from the strain-induced pseudo-magnetic field.

\section*{acknowledgments}
D.-N.Ly and N.-H.Phan are funded by Ho Chi Minh City University of Education Foundation for Science and Technology under grant numbers CS.2019.19.43TD and CS.2019.19.44TD. This paper is funded by Foundation for Science and Technology of Vietnam Ministry of Education and Training under grant number B2022-SPS-09-VL. This paper was carried out by the high-performance cluster at Ho Chi Minh City University of Education, Vietnam.

\pagebreak
\onecolumngrid

\setcounter{equation}{0}
\setcounter{figure}{0}
\setcounter{table}{0}
\setcounter{page}{1}

\renewcommand{\thepage}{S-\arabic{page}} 
\renewcommand{\thesection}{S-\Roman{section}}  
\renewcommand{\thetable}{S-\Roman{table}}  
\renewcommand{\thefigure}{S-\arabic{figure}}
\renewcommand{\theequation}{S-\arabic{equation}}

\begin{center}
\textbf{\large
Supplemental materials for:\\Thermal effect on magnetoexciton energy spectra in monolayer \\transition-metal dichalcogenides
}\\

Duy-Nhat Ly $^{1,\dagger}$, Dai-Nam Le $^{2,\dagger}$, Ngoc-Hung Phan $^{1}$ and Van-Hoang Le $^{1}$\\
\textit{$^1$Computational Physics Key Laboratory K002, Department of Physics, Ho Chi Minh City University of Education, Ho Chi Minh City 72759, Vietnam\\
$^2$ Department of Physics, University of South Florida, Tampa, Florida 33620, USA\\
$^{\dagger}$These authors contributed equally to this work.
}\\

Emails: nhatld@hcmue.edu.vn (D.-N. Ly), dainamle@usf.edu (D.-N. Le), hoanglv@hcmue.edu.vn (V.-H. Le).\\

Links: \url{http://link.aps.org/supplemental/
10.1103/PhysRevB.107.155410}.

\end{center}

\maketitle
\tableofcontents

\section{\label{sec:S1new} Separation of center-of-mass motion}
\label{Separation}

\subsection{\label{S1.1} Hamiltonian for an electron-hole system in a magnetic field }
The Hamiltonian for an electron-hole system in a magnetic field can be written as
\begin{equation}\label{Sequ9}
   {\hat H_{ex}} = \frac{1}{{2{m^*_e}}}{\hat {\mathbf p_e}}^2 + \frac{1}{{2{m^*_h}}}{\hat {\mathbf p_h}}^2 + \frac{{eB}}{{2{m^*_e}}}{\hat l^e_{z}} - \frac{{eB}}{{2{m^*_h}}}{\hat l^h_{z}} + \frac{{{e^2}{B^2}}}{{8{m^*_e}}}r_e^2 + \frac{{{e^2}{B^2}}}{{8{m^*_h}}}r_h^2+ {\hat V}_{h-e}\left( {|{{\mathbf r}_h} - {{\mathbf r}_e}|} \right),
\end{equation}
where $m^*_e$ and $m^*_h$ are the effective masses of the electron and hole, respectively. To separate the variables, we use the center of mass (c.m.) coordinates $\mathbf R =X\, \mathbf i + Y \,\mathbf j $ and the relative coordinates $\mathbf r = x\, \mathbf i + y \,\mathbf j $ defined by the transformation
\begin{eqnarray}\label{Sequ14}
X &=& \frac{{{m^*_h}}}{{{m^*_h} + {m^*_e}}}{x_h} + \frac{{{m^*_e}}}{{{m^*_h} + {m^*_e}}}{x_e}, \qquad
x = {x_e} - {x_h}, \nonumber\\
Y &=& \frac{{{m^*_h}}}{{{m^*_h} + {m^*_e}}}{y_h} + \frac{{{m^*_e}}}{{{m^*_h} + {m^*_e}}}{y_e}, \qquad
y = {y_e} - {y_h}.
\end{eqnarray}

For the separation, we need to rewrite all the terms in Hamiltonian \eqref{Sequ9} in the coordinates $ \left( {\mathbf r,\mathbf R} \right)$ and, as a result, obtain the following
\begin{eqnarray}\label{Sequ18}
\frac{1}{{2{m^*_e}}}{{\hat {\mathbf p}}_e}^2 + \frac{1}{{2{m^*_h}}}{{\hat {\mathbf p}}_h}^2 
= \frac{1}{{2M}}{{\hat {\mathbf P}}^2} + \frac{1}{{2\mu }}{{\hat {\mathbf p}}^2},\qquad\qquad\qquad\\
\label{Sequ18b}
\frac{eB}{2m^*_e}{\hat l^e_{z}} - \frac{eB}{2m^*_h}{\hat l^h_{z}} 
=\frac{1-\sigma}{1+\sigma}\,\frac{{eB}}{2\mu}\,{\hat l}_z 
+\frac{e}{{2M}}( {\mathbf B \times \mathbf r}).\hat {\mathbf P}
 + \frac{e}{{2\mu }}( {\mathbf B \times \mathbf R}) \cdot \hat {\mathbf p}\, \,,\\
\label{Sequ18c}
\frac{1}{m^*_e} r_e^2 + \frac{1}{m^*_h}r_h^2 
= \frac{1}{\mu }{R^2} + \frac{1-\sigma}{(1+\sigma)\mu}\, 2\mathbf r \cdot \mathbf R + \left(\frac{1}{\mu }
     - \frac{3}{M}\right) {r^2}.\qquad
\end{eqnarray}
Here, the total mass $M$, the reduced mass $\mu$, and the ratio of masses $\sigma$ are defined as
\begin{equation}\label{Sequ19}
M = {m^*_h} + {m^*_e},\qquad \frac{1}{\mu } = \frac{1}{{{m^*_e}}} + \frac{1}{{{m^*_h}}},\qquad \sigma=m^*_e/m^*_h;
\end{equation}
$\hat {\mathbf P}$ is the c.m. momentum; $\hat {\mathbf p}$ is the momentum of the relative motion between the electron and hole; $\hat l_z=x {\hat p_y}-y{\hat p_x}$ is the angular momentum of the relative motion on Oxy plane.

Using equations \eqref{Sequ18}, \eqref{Sequ18b}, and \eqref{Sequ18c}, we can rewrite the Hamiltonian in the coordinate system of the c.m. as
\begin{eqnarray}\label{Sequ28}
{\hat H}_{ex} &=& \frac{1}{2\mu }{{\hat {\mathbf p}}^2} +\frac{1-\sigma}{1+\sigma}\,\frac{{eB}}{2\mu}\,{\hat l}_z 
              + \frac{M-3\mu}{M \mu} \,\frac{{e^2}{B^2}}{8}{r^2}  + {\hat V_{h-e}}(r) \nonumber\\
&&+\frac{1}{2M} {{\hat {\mathbf P}}^2} + \frac{{e^2}{B^2}}{8\mu} {R^2} 
+ \frac{{e^2}{B^2}}{4\mu} \frac{1-\sigma}{1+\sigma} \,\mathbf r \cdot \mathbf R \nonumber \\
&&+\frac{e}{{2M}}( {\mathbf B \times \mathbf r}) \cdot \hat {\mathbf P}
 + \frac{e}{{2\mu }}( {\mathbf B \times \mathbf R}) \cdot \hat {\mathbf p}\,\,.
\end{eqnarray}

\subsection{\label{S1.2} Pseudomomentum of the electron-hole system in a magnetic field }

At first sight, Hamiltonian (\ref{Sequ28}) is not variable-separable for the coordinates $\mathbf R$ and $\mathbf r$ because the total momentum $\hat {\mathbf P}$ is not conserved. However, there is another conservative quantity for the considered system \cite{Scomment2020, SAvron1978, SRuder1994}, called pseudomomentum, defined as
\begin{equation}\label{Sequ35}
{\hat {\mathbf P}_0} = \hat {\mathbf P} - \frac{1}{2}e\mathbf B \times \mathbf r\,\,.
\end{equation}

We will insert this vector in the Hamiltonian instead of $\mathbf P$. For this purpose, we first calculate the quadratic form $\hat {\mathbf P}_0^2$ and then express the kinetic energy operator of c.m. via $\hat {\mathbf P}_0$ as
\begin{equation}\label{Sequ38}
\quad \frac{1}{{2M}}{\hat {\mathbf P}^2} = \frac{1}{{2M}}\hat {\mathbf P}_0^2 + \frac{e}{{2M}}\left( {\mathbf B \times \mathbf r} \right) \cdot \hat {\mathbf P}_0 + \frac{{{e^2}{B^2}}}{{8M}}{r^2}.
\end{equation}
The other term in the Hamiltonian containing the total momentum $\hat {\mathbf P}$  can be expressed via $\hat {\mathbf P}_0$ too,
\begin{eqnarray}\label{Sequ40}
\frac{e}{{2M}}( {\mathbf B \times \mathbf r}) \cdot \hat {\mathbf P} 
= \frac{e}{{2M}}( {\mathbf B \times \mathbf r}) \cdot {{\hat {\mathbf P}}_0} + \frac{{{e^2}{B^2}}}{{4M}}{r^2}.
\end{eqnarray}
Using equations (\ref{Sequ38}) and (\ref{Sequ40}), we rewrite the Hamiltonian (\ref{Sequ28}) into
\begin{eqnarray}\label{Sequ42}
{{\hat H}_{ex}} 
&=& \frac{1}{{2\mu }}{{\hat {\mathbf p}}^2} + \frac{1-\sigma}{1+\sigma}\frac{{eB}}{2\mu}{{\hat l}_z} + \frac{{{e^2}{B^2}}}{{8\mu }}{r^2} + {\hat V_{h-e}}\left( r \right) \nonumber\\
&+& \frac{1}{{2M}}\hat {\mathbf P}_0^2 + \frac{e}{M}( {\mathbf B \times \mathbf r}).{{\hat {\mathbf P}}_0} + \frac{{{e^2}{B^2}}}{{8\mu }}{R^2} + \frac{1-\sigma}{1+\sigma}\frac{{{e^2}{B^2}}}{4\mu}\mathbf r.\mathbf R\, 
+ \frac{e}{{2\mu }}( {\mathbf B \times \mathbf R} ).\hat {\mathbf p}\,\,,
\end{eqnarray}
which now contains ${\mathbf P}_0$ only. 

We can also confirm that the pseudomomentum ${\hat {\mathbf P}_0}$ commutes with the Hamiltonian (\ref{Sequ42}) by calculating commutators of ${\hat {\mathbf P}_0}$ with all the Hamiltonian terms. The following commutator relations
\begin{equation}\label{Sequ47b}
\left[{\hat {\mathbf P}}_0\,\,,  \frac{{{e^2}{B^2}}}{{8\mu }}{r^2}\right] = 0\,\,,
 \quad \left[{\hat {\mathbf P}}_0, {\hat V_{h-e}}(r)\right] =0\,\,,
 \quad \left[{\hat {\mathbf P}}_0,  \frac{e}{M}( {\mathbf B \times \mathbf r}).{{\hat {\mathbf P}}_0}\right] = 0
\end{equation}
are trivially obtained because the operator $\hat {\mathbf P}_0$ does not contain the differential operator respecting $\mathbf r$.
For all other terms in the Hamiltonian, we can calculate and receive the following commutators
\begin{eqnarray}\label{Sequ45a}
&&\left[ {\hat {\mathbf P}}_0,  {{\hat {\mathbf p}}^2} \right] =  - i\hbar\,e\mathbf B \times \hat {\mathbf p} \,\,,\quad
\left[{\hat {\mathbf P}}_0,  {\hat l_z} \right] =\frac{1}{2}  i\hbar\, eB \,{\mathbf r}\,\,,\quad
\left[{\hat {\mathbf P}}_0, {R^2}\right]  = - \frac{1}{2} i\hbar\, {\mathbf R}\,\,,\nonumber\\
&&\left[{\hat {\mathbf P}}_0,  \mathbf r.\mathbf R \right] =  -  i\hbar \,{\mathbf r}\,\,,\quad\quad
\left[{\hat {\mathbf P}}_0,  ( {e\mathbf B \times \mathbf R} ).\hat {\mathbf p} \right] =  i\hbar\, e{\mathbf B} \times {\vec {\mathbf p}} + \frac{1}{2} i\hbar\,{e^2}{B^2}\, {\mathbf R}\,\,.
\end{eqnarray}
From equations (\ref{Sequ47b}) and (\ref{Sequ45a}), we can prove
\begin{equation} \label{Sequ51}
\left[ {{{\hat {\mathbf P}}_0},\hat H_{ex}} \right] = 0\,\,,
\end{equation}
meaning that $\hat {\mathbf P}_0$ is an integral of motion, and $\hat {\mathbf P}_0$ and $\hat H_{ex}$ have mutual eigenfunctions.

\subsection{\label{S1.3} Variable-separation and the Hamiltonian for the relative motion}

We will use eigenfunctions of the pseudomomentum ${\hat {\mathbf P}_0}$ to separate the variables in the total Hamiltonian \eqref{Sequ28}.
For the first move, it is easy to verify that the wave function
\begin{equation} \label{Sequ59}
\Psi \left({\mathbf R}, {\mathbf r} \right) = \exp \left\{ \frac{i}{\hbar } \left( {\mathbf K} 
           + \frac{e}{2}{\mathbf B} \times {\mathbf r} \right).{\mathbf R} \right\}\psi (\mathbf r)
\end{equation}
is an eigenfunction of the operator ${\hat {\mathbf P}_0}$ with an eigenvector $\mathbf K$, i. e., satisfies the eigenvalue equation  
\begin{equation} \label{Sequ52}
{\hat {\mathbf P}_0}\Psi \left( {\mathbf R,\mathbf r} \right) = \mathbf K\Psi \left( {\mathbf R,\mathbf r} \right).
\end{equation}
Because of the commutation relation $\left[\hat {\mathbf P}_0,\hat H_{ex} \right] =0$, we can choose $\psi (\mathbf r)$ so that the wave function $\Psi \left({\mathbf R}, {\mathbf r} \right)$ becomes an eigenfunction of the Hamiltonian $\hat H_{ex}$, i. e.,
\begin{equation} \label{Sequ53}
{\hat H_{ex}}\Psi \left( {\mathbf R,\mathbf r} \right) = E_{total} \Psi \left( {\mathbf R,\mathbf r} \right).
\end{equation}
Denoting $\hat U = \exp \left\{ \frac{i}{\hbar } \left({\mathbf K} + \frac{e}{2}{\mathbf B} \times {\mathbf r} \right) \cdot {\mathbf R} \right\}$, we can rewrite equation \eqref{Sequ53} into
 \begin{eqnarray} \label{Sequ54}
{\hat H_{ex}}{\hat U}\psi \left({\mathbf r} \right) = E_{total} \,{\hat U}\psi \left({\mathbf r} \right)
\rightarrow {\hat U}^{-1}{\hat H_{ex}}{\hat U}\psi \left({\mathbf r} \right) = E_{total} \psi \left({\mathbf r} \right).
\end{eqnarray}
The function $\psi (\mathbf r)$, depending only on $\mathbf r$, can be considered as a wave function of the electron-hole relative motion. Consequently, equation \eqref{Sequ54} (subtracted by the pseudokinetic energy of the exciton c.m.) is the Schr{\"o}dinger equation for relative motion with the Hamiltonian defined by the transformation
\begin{equation}\label{Sequ55}
{\hat H}_{rel}={\hat U}^{-1} {\hat H}_{ex} {\hat U}-\frac{1}{2M}K^2
\end{equation}
with the relative energy $E=E_{total}-\frac{1}{2M}K^2$.

We can directly calculate all the terms of the Hamiltonian $\hat H_{rel}$ with the following results
\begin{eqnarray}\label{Sequ69}
&{\hat U}^{ - 1}{{\hat {\mathbf p}}^2}\,{\hat U} = {\hat {\mathbf p}^2}
 - \left( {e {\mathbf B} \times {\mathbf R}} \right)\cdot \hat {\mathbf p} 
       + \frac{1}{4}{{e^2}{B^2}}{R^2}\,\,,\nonumber\\
& {\hat U}^{ - 1} ( e{\mathbf B \times \mathbf R} )\cdot \hat {\mathbf p}\,{\hat U}  =  - \frac{1}{2}{e^2 B^2}{R^2} 
+ (e{\mathbf B} \times {\mathbf R}) \cdot \hat {\mathbf p}\,\,,\nonumber\\
&{\hat U}^{ - 1} \, {\hat l_z} {\hat U} 
=  {{\hat l}_z}- \frac{1}{2}\,{e}{B}\,{\mathbf r}\cdot {\mathbf R}\,\,,\quad
{\hat U}^{ - 1}\hat {\mathbf P}_0^2\, {\hat U} 
      = {K^2}\,\,,\quad
{\hat U}^{ - 1} ( e{\mathbf B \times \mathbf r}) \cdot {{\hat {\mathbf P}}_0}\, {\hat U} 
      =  \left( {e{\mathbf B} \times \mathbf K} \right) \cdot {\mathbf r}\,\, ,\nonumber\\
&{\hat U}^{ - 1}{r^2}\,{\hat U} =  {r^2}\,\,,\qquad
{\hat U}^{ - 1}{\hat V_{h-e}}(r)\, {\hat U} = {\hat V_{h-e}}(r)\,\,,\qquad {\hat U}^{ - 1}{R^2}\, {\hat U} ={R^2},
\qquad {\hat U}^{ - 1}{\mathbf r} \cdot {\mathbf R} \,{\hat U} 
    = \,{\mathbf r} \cdot {\mathbf R}\,\,.
\end{eqnarray}
From equations \eqref{Sequ55} and \eqref{Sequ69}, we obtain the Hamiltonian for the relative motion between the electron and the hole as
\begin{eqnarray}\label{Sequ78}
{{\hat H}_{rel}} 
&=& \frac{1}{{2\mu }}{{\hat {\mathbf p}}^2} +\frac{1-\sigma}{1+\sigma}\,\frac{{eB}}{2\mu}\,{\hat l}_z +
\frac{{{e^2}{B^2}}}{{8\mu }}{r^2} + {\hat V_{h-e}}\left( r \right) 
- \frac{e}{M}\left( {\mathbf B \times \mathbf K} \right)\cdot \mathbf r,
\end{eqnarray}
which is the Hamiltonian in equation (2) of the main text.

\section{\label{sec:S2new} Thermo-induced motional Stark potential}

In equation \eqref{Sequ78}, we now consider the term 
\begin{equation}\label{Sequ1-2}
V_{mS} (\mathbf{r})=- \frac{e}{M}\left( {\mathbf B \times \mathbf K} \right) \cdot \mathbf r\,\,,
\end{equation}
which was neglected in the previous works for the exciton energies in monolayer TMDCs. The term \eqref{Sequ1-2} contains the exciton c.m. pseudomomentum $\mathbf{K}$ that inspires us to establish its relation to the thermal motion of excitons in a monolayer TMDC. 

Excitons in a monolayer TMDCT can be regarded as a gas of thermal motion with kinetic energy $K^2/2M$. We first compare the average distance between two excitons $\lambda_{ex}=1/\sqrt{n_X}$ with the thermal wavelength $\lambda_{th} = \sqrt{{2\pi\hbar^2}/{(M k_B T)}}$ to justify which statistics should be applied for the kinetic energy distribution. Here, $n_X$ is the exciton density -- the number of neutral excitons per unit of area; $k_B$ is the Boltzmann constant. For the Maxwell-Boltzmann statistics to be applied, the thermal wavelength should be much smaller than the average exciton distance that leads to the limit for the exciton density as
\begin{equation}\label{Sequ2-2}
\lambda_{th} \ll \lambda_{ex}\rightarrow \quad n_X \ll n_{max}=\frac{M k_B T}{2\pi\hbar^2}\,\,.
\end{equation}
For room temperature, 300K, we calculate the limit for the exciton density in monolayer $\text{WSe}_2$ where $M=0.8 \times m_e$ and receive a result $n_{max-300\text{K}}=4.3\times 10^{12} \text{cm}^{-2}$. On another side, the exciton density is modulated around $10^{11} \text{--} 10^{12} \text{ cm}^{-2}$ in real experiments for monolayer WSe$_2$ \cite{SPoellmann2015, SYan2015, SYou2015, SSteinleitner2017}, which is much smaller than its limit  $n_{max-300\text{K}}$ at room temperature. Therefore, we can estimate the root-mean-square of the c.m. pseudomomentum at temperature $T$ by the equipartition theorem as $\frac{1}{2M}\overline{ K^2} =  k_B T$. 

In exciton gas, pseudomomentum $K$ has an assemply of values $(K_1, K_2, ...)$ obeyed the Maxwell-Boltzmann distribution. To obtain the average energy $\overline {E}$ of different values $K$, we principally need to calculate energies for each value $K_j$ and then average the calculated energies. However, we can do this in another way by  calculating energy at the average pseudomomentum $\overline K=\sqrt{\overline{K^2}}$. The results must be the same $E(\overline{K}) \simeq \overline{E}$. The base for this assumption is the perturbation theory calculation that the thermo-induced potential leads to a small correction to the energy, which is proportional to the quantity $K^2$ as $E=E^{(0)}+b\times K^2$ so that $\overline{E}=E^{(0)}+b\times \overline{K^2}$. We also verify numerically that the two average methods (the average energy by different pseudomomentum vs. the energy at the average pseudomomentum) give the same results presented in Table \ref{tabS1a} with diminutive differences of about 0.1 meV, except for $3s$-state with about 0.5 meV difference (relative error of 0.68\%).

Consequently, from the above discussion, we can estimate the average pseudomomentum of exciton c.m. motion via temperature and rewrite the motional stark potential in the form 
\begin{eqnarray}\label{Seq2-3}
V_{mS} (\mathbf{r}) =- \sqrt{\frac{2k_B T}{M}} eB\, x \,\,.
\end{eqnarray}
Here, without losing generality, we consider the case where $\mathbf{K}$ is along the $Oy$ axis so that vector $\mathbf{B} \times \mathbf{K}$ is along the $Ox$ axis.

For the system of an atom in external electric and magnetic field, the consideration of the term \eqref{Sequ78} leads to additional contribution in the Stark effect although being originated from the magnetic field; thus called the motional Stark effect \cite{SFarrelly1994, SSchweiner2017}. However, because this term is related to the thermal motion as we shown via the formula \eqref{Seq2-3}, we call it the thermo-induced motional Stark potential. 

\begin{figure}[H]
\begin{center}
\includegraphics[width=0.4 \columnwidth]{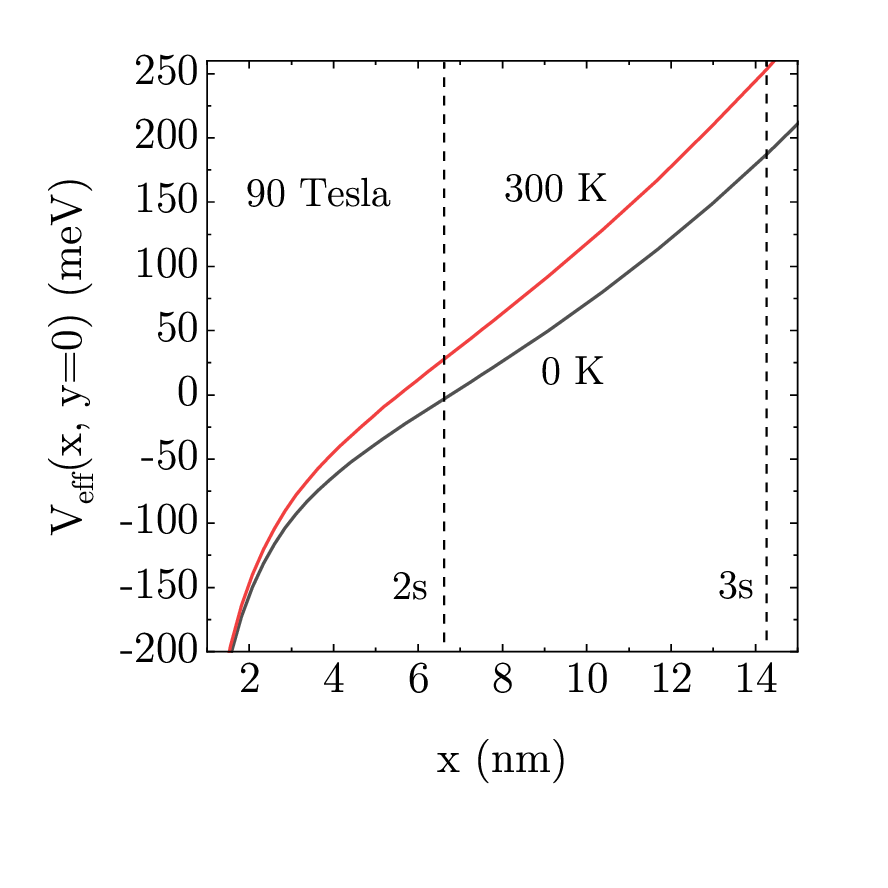}
\caption{The effective potentials for relative motion in two cases: $T=0K$ (black) and $T=300K$ (red), i.e., without and with the motional Stark term. The difference between the potentials is noticeable for $2s$ and higher Rydberg states, where the exciton radius is larger than 6.5 nm.}
\label{figS1}
\end{center}
\end{figure}

The effective potential in Hamiltonian \eqref{Sequ78} consists of three terms as ${\hat V}_{eff}={\hat V}_{diamag}+ {\hat V}_{h-e}+{\hat V}_{mS}$, where the diamagnetic term $V_{diamag.} (\mathbf{r})=\frac{{e^2} {B^2}}{8 \mu } {r}^2$ is quadratic proportional to the electron-hole distance and consequently dominant at the large separation of the electro-hole pair compared to the thermo-induced motional Stark term $\hat V_{mS}$, which is linearly proportional to the variable $x$. As a result, there is no tunneling effect, even considering the thermo-induced motional Stark potential. Instead of tunneling, we expect the thermo-induced motional Stark potential causes the Stark shift in the energy spectra.This fact is also demonstrated clearly in Fig. \ref{figS1}, where the inclusion of the thermal effect causes a higher raising in the effective potential. 

\begin{center}
	\begin{table}[H]
		\caption{\label{tabS1a}Exciton energies (meV) in the magnetic field of 90 Tesla with different values of $K^2$ calculated for monolayer $\text{WSe}_2$ with $r_0=4.2086$ nm, $\mu=0.2039\, m_e$, $\kappa=4.5$\,\,. It also indicates energies averaged by $K^2$ and calculated at the average value of $K^2$. The energy differences between the two different calculation methods are in the last row.  }
		\begin{ruledtabular}
			\begin{tabular}{r r r r r r r r r r r}
$K$ ($\hbar/a_0$)	&	$K^2$	($\hbar^2/a^2_0 $)        &	                   &  &	Energy (meV)  &	 &	\\
      $\times 10^{-3} $  &	$\times 10^{-3} $    &	$1s$	& $2p_{-1}$ &	$2p_{+1}$ &	 $2s$ 	&	$3s$ 	\\
\hline
0	&	0	&	-165.807&	-33.280&	-31.645	& -6.750	 & 68.124  \\
5  &	0.025	&	-165.808 &	-33.292 &	-31.657 &	-6.743	& 68.148 \\
10	&	0.100	&	-165.810 &	-33.329	& -31.693	& -6.721	& 68.221\\
15	&	0.225	&	-165.813 &	-33.390 & -31.752 &	-6.684	& 68.339\\
20	&	0.400	&	-165.818 &	-33.479 &	-31.831&	-6.634	& 68.501 \\
25	&	0.625	&-165.824&	-33.597&	-31.927&	-6.570&	68.700\\
30	&	0.900	&	-165.831& 	-33.748&	-32.037&	-6.495	& 68.934\\
35	&	1.225	&	-165.839&	-33.934&	-32.157&	-6.408&	69.195\\
40	&	1.600	&	-165.849&	-34.158&	-32.285&	-6.312&	69.478\\
45	&	2.025	&	-165.860&	-34.420&	-32.417&	-6.207	& 69.777\\
50	&	2.500	&	-165.872&	-34.722&	-32.552&	-6.094&	70.086\\
55	&	3.025	&	-165.886	 & -35.062	& -32.690& 	-5.977	 & 70.397\\
60	&	3.600	&	-165.901	& -35.439&	-32.831	&-5.855&	70.706\\
65	&	4.225	&-165.917&	-35.852	&-32.976&	-5.732& 	71.004\\
70	&	4.900	&-165.935&	-36.299&	-33.125&	-5.608&	71.284\\
75	&5.625	&	-165.954 & 	-36.779& 	-33.280	&-5.485&	71.538\\
80	&	6.400	&-165.974&	-37.288&	-33.440	&-5.365	&71.757\\
85	&	7.225	&-165.995&	-37.827&	-33.608&	-5.251&	71.928\\
90	&	8.100	&-166.018	&-38.394&	-33.782&	-5.144&	72.036\\
95	&	9.025	&-166.042&	-38.987&	-33.965&	-5.046	&72.060\\
100	&	10.000	&-166.068&	-39.605&	-34.156&	-4.961&	71.977\\
\hline
Average	&	$3.417 $&	-165.896&	-35.375	&	-32.657	&	-6.002	&	70.104	\\
\hline
& $E (\overline{K^2}=3.417) 	$& -165.895 &-35.284 & -32.774 &-5.904 & 70.580\\
\hline
   &Difference (meV)&0.001 & 0.091 & 0.117 & 0.098 & 0.479  \\
			\end{tabular}
		\end{ruledtabular}
	\end{table}
\end{center}

\section{\label{sec:S1}Feranchuk-Komarov operator method for magnetoexciton energies under influence of thermo-induced Stark potential}
\label{FKMethod}
\subsection{\label{sec:S1A}Schr\"odinger equation in the formalism of creation and annihilation operators}

Magnetoexciton under the influence of thermo-induced Stark potential is described by the Schr\"odinger equation (2) in the main text. We rewrite it in a dimensionless form where the energy $E$ and the coordinates $x, y$ are given in the effective Hartree unit $2 Ry^{*}=\mu e^4/16\pi^2\varepsilon_0^2\hbar^2$ and the effective Bohr radius $a_0^{*}=4\pi\varepsilon_0\hbar^2/\mu e^2$, respectively; the dimensionless magnetic intensity $\gamma$ is related to the magnetic field by the equation $B=2\gamma \times \mu\hbar Ry^{*}/e$. Then the Levi-Civita transformation \cite{Sgiang1993}
\begin{equation}
\label{Seq11}
x=u^2-v^2,\; y=2uv
\end{equation}
is applied to the dimensionless Schr{\"o}dinger equation to rewrite it in the $(u, v)$ space as 
\begin{eqnarray}\label{Seqn5}
\left\{ { - \frac{1}{8}\left( {\frac{{{\partial ^2}}}{{\partial {u^2}}} + \frac{{{\partial ^2}}}{{\partial {v^2}}}} \right) + \left( {\frac{{1 - \sigma }}{{1 + \sigma }}\frac{\gamma }{2}{\mkern 1mu} {{\hat l}_z} - E} \right)\left( {{u^2} + {v^2}} \right)} \right. + \frac{{{\gamma ^2}}}{8}{\left( {{u^2} + {v^2}} \right)^3} \nonumber \\
\left. {+ \frac{{\sqrt {2\sigma {\mkern 1mu} \theta } }}{{1 + \sigma }}\gamma \left( {{u^4} - {v^4}} \right) + {\mkern 1mu} V(u,v)} \right\}\psi (u,v) = 0,
\end{eqnarray}
where $\theta=k_B T/ 2 Ry^{*}$ is dimensionless temperature; and $V (u, v)=\left( {u}^2+{v}^2\right) V_{RK}$. Here, from now on, we use the notation $V_{RK}$ for the interaction potential of the electron and hole $\hat V_{h-e}$. Also, for convenience in calculations, we rewrite the potential $V (u, v)$ in the Laplace transform \cite{Sprudnikov1992} as
\begin{equation}\label{Seq13}
V(u,v)=- \frac{1}{ \kappa} 
\int\limits_{0}^{+\infty} \frac{dq}{ \sqrt{1+\alpha^2 q^2} }\;
\textrm{e}^{-q(u^2+v^2)}(u^2+v^2),
\end{equation}
where the dimensionless parameter $\alpha=r_0/\kappa a^*_0$ is used instead of the screening length $r_0$. In equations \eqref{Seqn5} and \eqref{Seq13}, we have used the formulas 
\begin{equation}\label{Seq15}
r=u^2+v^2,\quad\quad
{\hat l_z}= -\frac{i}{2}\left( v\frac{\partial}{\partial u} - u\frac{\partial}{\partial v} \right)
\end{equation}
obtained by the Levi-Civita transformation \eqref{Seq11} for distance $r$ and angular momentum $\hat l_z$.

Equation \eqref{Seqn5} is similar to those describing a two-dimensional anharmonic oscillator, suggesting we use the harmonic oscillator wave functions as a basis set and calculate necessary matrix elements by  the algebraic approach based on annihilation and creation operators, well-known as the Feranchuk-Komarov operator method in Refs.~\cite{SHoangbook2015, SNguyen2019}. The first step is to rewrite the equation in the presentation of annihilation and creation operators, defined as
\begin{eqnarray}\label{Seqn6}
\hat a = \frac{1}{{\sqrt 2 }}\left( {\hat \alpha  - i\hat \beta } \right),
\quad \hat a^ + = \frac{1}{{\sqrt 2 }}\left( {{{\hat \alpha }^ + } + i{{\hat \beta }^ + }} \right),
\quad \hat b = \frac{1}{{\sqrt 2 }}\left( {\hat \alpha  + i\hat \beta } \right),
\quad \hat b^ + = \frac{1}{{\sqrt 2 }}\left( {{{\hat \alpha }^ + } - i{{\hat \beta }^ + }} \right),
\end{eqnarray}
where
\begin{eqnarray}\label{Seqn7}
\hat \alpha  = \sqrt {\frac{\omega }{2}} \left( {u + \frac{1}{\omega }\frac{\partial }{{\partial u}}} \right),
\quad \hat \alpha ^ +  = \sqrt {\frac{\omega }{2}} \left( {u - \frac{1}{\omega }\frac{\partial }{{\partial u}}} \right), \nonumber \\
\quad \hat \beta  = \sqrt {\frac{\omega }{2}} \left( {v + \frac{1}{\omega }\frac{\partial }{{\partial v}}} \right),
\quad \hat \beta ^ +  = \sqrt {\frac{\omega }{2}} \left( {v - \frac{1}{\omega }\frac{\partial }{{\partial v}}} \right).
\end{eqnarray}
Here, $\omega$ is a free parameter that manipulates the convergence rate of numerical calculations and does not affect the final energy spectrum. This parameter has been introduced and discussed in many references such as \cite{SHoangbook2015, SNguyen2019} and references therein. These operators must obey the following commutation relations
\begin{eqnarray}\label{Seqn8}
\left[ {\hat a,{{\hat a}^ + }} \right] = \left[ {\hat b,{{\hat b}^ + }} \right] = 1,
\quad \left[ {\hat a,\hat a} \right] = \left[ {{{\hat a}^ + },{{\hat a}^ + }} \right] = \left[ {\hat b,\hat b} \right] = \left[ {{{\hat b}^ + },{{\hat b}^ + }} \right] = 0,
\end{eqnarray}
typical for annihilation and creation operators.

All the operators of Eq.~\eqref{Seqn5} can be presented in terms of creation and annihilation operators as
\begin{eqnarray}\label{Seqn9}
&&-\left( \frac{{{\partial ^2}}}{{\partial {u^2}}} + \frac{{{\partial ^2}}}{{\partial {v^2}}} \right)
= \omega \left( {-\hat a\hat b - {{\hat a}^ + }{{\hat b}^ + } + {{\hat a}^ + }\hat a + {{\hat b}^ + }\hat b + 1} \right) \equiv \omega \hat{T}, \nonumber \\
&&{{\hat l}_z} = \frac{1}{2}\left( {{{\hat a}^ + }\hat a - {{\hat b}^ + }\hat b} \right),\nonumber\\
&&{u^2} + {v^2} = \frac{1}{\omega }\left( {\hat a\hat b + {{\hat a}^ + }{{\hat b}^ + } + {{\hat a}^ + }\hat a + {{\hat b}^ + }\hat b + 1} \right), \nonumber \\
&&{u^4} - {v^4} = \left( {{{\hat a}^ + }{{\hat a}^ + } + {{\hat b}^ + }{{\hat b}^ + } + \hat a\hat a + \hat b\hat b + 2{{\hat b}^ + }\hat a + 2{{\hat a}^ + }\hat b} \right)\left( {\hat a\hat b + {{\hat a}^ + }{{\hat b}^ + } + {{\hat a}^ + }\hat a + {{\hat b}^ + }\hat b + 1} \right).
\end{eqnarray}
Therefore, Eq.~\eqref{Seqn5} is expressed as a secular equation
\begin{eqnarray}\label{Seqn10}
\hat H\left| \psi  \right\rangle  - E\,\hat R\left| \psi  \right\rangle  = 0,
\end{eqnarray}
in which the explicit expressions of $\hat{H}$ and $\hat{R}$ are given as
\begin{eqnarray}\label{Seqn11}
&&\hat R = {{\hat a}^ + }\hat a + {{\hat b}^ + }\hat b + 1 + \hat a\hat b + {{\hat a}^ + }{{\hat b}^ + },  \\
&&\hat H = \frac{{{\omega ^2}}}{8} \hat{T} + \frac{{1 - \sigma }}{{1 + \sigma }}\frac{\gamma }{2}\hat{R}\, \hat{l}_z + \frac{{{\gamma ^2}}}{{8{\omega ^2}}}{\hat{R}^3} - \omega \,{{\hat V}} + \frac{{\sqrt {2\sigma \theta} }}{{1 + \sigma }}\frac{\gamma }{{2\omega }}{{\hat V}_{mS}}.
\end{eqnarray}

In Eq.~\eqref{Seqn11}, operator ${{\hat{V}}}$ describes the Rytova-Keldysh interaction in the creation-annihilation-operator formalism and is written as the following integral
\begin{eqnarray}\label{Seqn12}
{\hat V} =  - \frac{1}{\kappa }\int\limits_0^\infty  {\frac{{dq}}{{\sqrt {1 + {\alpha ^2}{\omega ^2}{q^2}} }}} \;{e^{ - q\hat R}}\hat R,
\end{eqnarray}
while operator 
\begin{eqnarray}\label{Seqn13}
{\hat V_{mS}} = \left( {{{\hat a}^ + }{{\hat a}^ + } + {{\hat b}^ + }{{\hat b}^ + } + \hat a\hat a + \hat b\hat b + 2{{\hat b}^ + }\hat a + 2{{\hat a}^ + }\hat b} \right)\left( {{{\hat a}^ + }\hat a + {{\hat b}^ + }\hat b + 1 + \hat a\hat b + {{\hat a}^ + }{{\hat b}^ + }} \right)
\end{eqnarray} 
is of the thermo-induced motional Stark potential.

\subsection{\label{sec:S1B}Matrix elements}

The basis set in this method is chosen by the wave functions of the two-dimensional harmonic oscillator as
\begin{eqnarray}\label{Seqn14}
\left| {k,m}, \omega \right\rangle  = \frac{1}{{\sqrt {\left( {k + m} \right)!\left( {k - m} \right)!} }}{\left( {{{\hat a}^ + }} \right)^{k + m}}{\left( {{{\hat b}^ + }} \right)^{k - m}}\left| {0\left( \omega  \right)} \right\rangle ,
\end{eqnarray}
with the running numbers $k = \left| m \right|, |m|+1, |m|+2, \ldots$ and  $m=0,\, \pm 1,\, \pm 2,\, \dots$. We note that the frequency $\omega$ can be regarded as a free parameter and will choose its value appropriate for regulating the convergence rate of the numerical solutions. Expanding the quantum wave function by this basis set as
\begin{equation}
\label{Seq30}
{| \psi^{(s)} (\omega)\rangle}=  \sum_{k=0}^{s} \sum_{m=0}^{k} C_{km}^{(s)} {| k, m, \omega\rangle}
\end{equation}
with $N=(s+1)(s+2)/2$ unknown coefficients $C_{km}^{(s)}$ needed to define and then putting these coefficients $C_{km}^{(s)}$ as a column matrix $\mathbb{C}$, Eq.~\eqref{Seqn10} turns into a usual form of a generalized secular equation of the eigenproblem of an $N \times N$-matrix $\mathbb{H}$ with respect to $\mathbb{R}$ as
\begin{eqnarray}\label{Seqn15}
\left( {\mathbb{H} - E\, \mathbb{R}} \right) \mathbb{C}= 0.
\end{eqnarray}
The matrix eigenvalue equation can be solved using subroutine dsygvx.f of the Linear Algebra Package (LAPACK) \cite{SLapack}. Here, it is noticed that Schr{\"o}dinger equation (2) is invariant under the transformation $x \rightarrow -x,\, \mathbf{B} \rightarrow - \mathbf{B}$; therefore, we can use only $m \geq 0$ in the expansion \eqref{Seq30}. In the wave function \eqref{Seq30}, we use only $(s+1)/(s+2)/2$ basis set functions so that the number $s$ can be regarded as an approximation order of the solutions. In practice, we will increment the $(s)$-order until getting the desired precision.

Now, the second step of the Feranchuk Komarov operator method is to calculate elements of the matrices $\mathbb{H}$ and $\mathbb{R}$. By algebraic calculations using the commutation relations \eqref{Seqn8}, we find out the following useful matrix elements
\begin{eqnarray}\label{Seqn16}
&&{N_{\scriptstyle k',m'\hfill\atop
\scriptstyle k,m\hfill}} = \langle k',m', \omega|\, \left( {{{\hat a}^ + }\hat a + {{\hat b}^ + }\hat b} \right)\, |k,m,\omega\rangle  = 2k \, {\delta _{k'k}}{\delta _{m'm}}, \nonumber \\
&&M_{\scriptstyle k',m'\hfill\atop
\scriptstyle k,m\hfill}^ +  = \langle k',m', \omega|\, {{\hat a}^ + }{{\hat b}^ + }\, |k,m,\omega\rangle  = \sqrt {{{(k + 1)}^2} - {m^2}} \, {\delta _{k',k + 1}}{\delta _{m'm}},\nonumber \\
&&{M_{\scriptstyle k',m'\hfill\atop
\scriptstyle k,m\hfill}} = \langle k',m', \omega|\, \hat a\hat b \, |k,m,\omega\rangle  = \sqrt {{k^2} - {m^2}} \, {\delta _{k',k - 1}}{\delta _{m'm}},
\end{eqnarray}
which help us to determine matrix elements as
\begin{eqnarray}\label{Seqn17}
{{\cal R}_{\scriptstyle k',m'\hfill\atop
\scriptstyle k,m\hfill}} &= \langle k',m', \omega| \hat{R} |k,m,\omega\rangle = & {N_{\scriptstyle k',m'\hfill\atop
\scriptstyle k,m\hfill}} + {M_{\scriptstyle k',m'\hfill\atop
\scriptstyle k,m\hfill}} + M_{\scriptstyle k',m'\hfill\atop
\scriptstyle k,m\hfill}^ + , \\
\mathcal{H}_{\scriptstyle k',m'\hfill\atop
\scriptstyle k,m\hfill} &= \langle k',m', \omega| \hat{H} |k,m,\omega\rangle = & {{\cal T}_{\scriptstyle k',m'\hfill\atop
\scriptstyle k,m\hfill}} + \frac{1-\sigma}{1+\sigma} \frac{\gamma m}{2} {\left( {{{\cal R}}} \right)_{\scriptstyle k',m'\hfill\atop
\scriptstyle k,m\hfill}} +  \frac{\gamma^2}{8 \omega^2} {\left( {{{\cal R}^3}} \right)_{\scriptstyle k',m'\hfill\atop
\scriptstyle k,m\hfill}} \nonumber\\
&& - \omega {\left( {{\mathcal{V}_{RK}}} \right)_{\scriptstyle k',m'\hfill\atop
\scriptstyle k,m\hfill}} + \frac{\sqrt{2\sigma \theta}}{1+\sigma} \frac{\gamma}{2 \omega} {\left( {{\mathcal{V}_{mS}}} \right)_{\scriptstyle k',m'\hfill\atop
\scriptstyle k,m\hfill}}  , \label{eqn17b}
\end{eqnarray}
where
\begin{eqnarray}
{{\cal T}_{\scriptstyle k',m'\hfill\atop
\scriptstyle k,m\hfill}} &&= \langle k',m', \omega| \hat{T} |k,m,\omega\rangle  = {N_{\scriptstyle k',m'\hfill\atop
\scriptstyle k,m\hfill}} - {M_{\scriptstyle k',m'\hfill\atop
\scriptstyle k,m\hfill}} - M_{\scriptstyle k',m'\hfill\atop
\scriptstyle k,m\hfill}^ + ,
\end{eqnarray}
and
\begin{eqnarray}\label{Seqn18}
{\left( {{{\cal R}^3}} \right)_{\scriptstyle k',m'\hfill\atop
\scriptstyle k,m\hfill}} && = \langle k',m', \omega| {{\hat R}^3}{\mkern 1mu} |k,m,\omega\rangle \nonumber\\
&&  = 2\left( {5{k^2} + 5k + 3 - 3{m^2}} \right)\left( {2k + 1} \right){\delta _{k'k}}{\delta _{m'm}} \nonumber \\
&& \quad + 6\left( {5{k^2} - 5k + 3 - 3{m^2}} \right)\sqrt {{k^2} - {m^2}} {\mkern 1mu} {\delta _{k',k - 1}}{\delta _{m'm}} \nonumber \\
&& \quad + 3\left( {2k - 1} \right)\sqrt {{k^2} - {m^2}} \sqrt {{{\left( {k - 1} \right)}^2} - {m^2}} {\delta _{k',k - 2}}{\delta _{m'm}} \nonumber \\
&& \quad + \sqrt {{k^2} - {m^2}} \sqrt {{{\left( {k - 1} \right)}^2} - {m^2}} \sqrt {{{\left( {k - 2} \right)}^2} - {m^2}} {\delta _{k',k - 3}}{\delta _{m'm}} \nonumber \\
&& \quad + 6\left( {5{k^2} + 5k + 3 - 3{m^2}} \right)\sqrt {{{\left( {k + 1} \right)}^2} - {m^2}}  {\delta _{k',k + 1}}{\delta _{m'm}} \nonumber \\
&& \quad + 3\left( {2k + 3} \right)\sqrt {{{\left( {k + 1} \right)}^2} - {m^2}} \sqrt {{{\left( {k + 2} \right)}^2} - {m^2}} {\delta _{k',k + 2}}{\delta _{m'm}} \nonumber \\
&& \quad + \sqrt {{{\left( {k + 1} \right)}^2} - {m^2}} \sqrt {{{\left( {k + 2} \right)}^2} - {m^2}} \sqrt {{{\left( {k + 3} \right)}^2} - {m^2}} {\delta _{k',k + 3}}{\delta _{m'm}}.
\end{eqnarray}
The matrix elements of the Rytova-Keldysh potential are calculated as
\begin{eqnarray}\label{Seqn19}
{\left( {{\mathcal{V}_{RK}}} \right)_{\scriptstyle k',m'\hfill\atop
\scriptstyle k,m\hfill}} && = \langle k',m', \omega|{\hat V}|k,m,\omega\rangle \nonumber\\
&& = \left[ {\left( {2k + 1} \right) {U_{k'k}} + \sqrt {{k^2} - {m^2}} {U_{k'k - 1}} + \sqrt {{{\left( {k + 1} \right)}^2} - {m^2}} {U_{k',k + 1}}} \right]{\delta _{m',m}},
\end{eqnarray}
where
\begin{eqnarray}\label{Seqn20}
U_{j'j} = && -\frac{1}{\kappa\,\alpha} \sum_{s=|m|}^{min(j,j')} \sum_{t=0}^{j'+j-2s} (-1)^{j'+j+t}
    {{j'+j-2s} \choose {t}} \nonumber\\
         && \times \sqrt{ {{j'+m}\choose{s+m}}} \sqrt{{{j'-m}\choose {s-m}}} 
               \sqrt{{{j+m}\choose {s+m}}} \sqrt{{{j-m}\choose {s-m}}}  \nonumber\\
&& \times \int\limits_{0}^{+\infty} \frac{dq}{(1+q)^{2s+t+1}\sqrt{q^2+1/\omega^2\alpha^2}}.
\end{eqnarray}
The matrix elements associated with the thermo-induced motional Stark effect are calculated as
\begin{eqnarray}\label{Seqn21}
{\left( {{\mathcal{V}_{mS}}} \right)_{\scriptstyle k',m'\hfill\atop
\scriptstyle k,m\hfill}} && = \langle k',m', \omega|{\hat V_{mS}}|k,m,\omega\rangle \nonumber\\
&& = \sqrt {\left( {k + m} \right)\left( {k + m - 1} \right)\left( {k + m - 2} \right)\left( {k - m} \right)} {\delta _{k' + m',k + m - 3}}{\delta _{k' - m',k - m - 1}} \nonumber\\
&& \quad + \sqrt {\left( {k - m + 3} \right)\left( {k - m + 2} \right)\left( {k - m + 1} \right)\left( {k + m + 1} \right)} {\delta _{k' + m',k + m + 1}}{\delta _{k' - m',k - m + 3}} \nonumber\\
&& \quad + \sqrt {\left( {k + m + 3} \right)\left( {k + m + 2} \right)\left( {k + m + 1} \right)\left( {k - m + 1} \right)} {\delta _{k' + m',k + m + 3}}{\delta _{k' - m',k - m + 1}} \nonumber\\
&& \quad + \sqrt {\left( {k - m} \right)\left( {k - m - 1} \right)\left( {k - m - 2} \right)\left( {k + m} \right)} {\delta _{k' + m',k + m - 1}}{\delta _{k' - m',k - m - 2}} \nonumber\\
&& \quad + \sqrt {\left( {k + m + 1} \right)\left( {k + m + 2} \right)} \left( {4k - 2m + 3} \right){\delta _{k' + m',k + m + 2}}{\delta _{k' - m',k - m}} \nonumber\\
&& \quad +3 \sqrt {\left( {k - m} \right)\left( {k + m + 1} \right)} \left( {2k + 1} \right){\delta _{k' + m',k + m + 1}}{\delta _{k' - m',k - m - 1}} \nonumber\\
&& \quad + \sqrt {\left( {k + m} \right)\left( {k + m - 1} \right)} \left( {2k - 2m + 1} \right){\delta _{k' + m',k + m - 2}}{\delta _{k' - m',k - m}} \nonumber\\
&& \quad + \sqrt {\left( {k - m} \right)\left( {k - m - 1} \right)} \left( {4k + 2m + 1} \right){\delta _{k' + m',k + m}}{\delta _{k' - m',k - m - 2}} \nonumber\\
&& \quad + \sqrt {\left( {k - m + 1} \right)\left( {k - m + 2} \right)} \left( {4k + 2m + 3} \right){\delta _{k' + m',k + m}}{\delta _{k' - m',k - m + 2}} \nonumber\\
&& \quad +3 \sqrt {\left( {k + m} \right)\left( {k - m + 1} \right)} {\delta _{k' + m',k + m - 1}}{\delta _{k' - m',k - m + 1}}.
\end{eqnarray}

\subsection{\label{sec:S1C}Exciton energy spectrum for magnetoexciton in $\text{WSe}_2$ monolayer at zero and room temperatures}

The third step is to solve Eq.~\eqref{Seqn10} with all the matrix elements calculated. The free parameter $\omega$ is chosen to make the convergence rate suitable for our computing resources and the required precision of 8 decimal places. This matrix eigenvalue equation can be solved by subroutine dsygvx.f of the Linear Algebra Package (LAPACK). We present here energies of magnetoexciton in monolayer $\text{WSe}_2$ for $1s$, $2s$, $2p_{-1}$, $2p_{+1}$, and $3s$ states  with magnetic intensity up to 500 Tesla. In the paper, we show fewer results for magnetic intensities, up to 90 Tesla produced in laboratories. We note that the method is applicable for the case with no magnetic field by dropping the thermo-induced term. 

\begin{center}
	\begin{table}[H]
		\caption{\label{tabS1}Energies of $1s$ states (meV) with different magnetic intensity $B$~(Tesla) at temperature 0~K and 300~K for $r_0$ = 4.2086~nm, $\mu$ = 0.2039~$m_e$, $\kappa = 4.5$.}
		\begin{ruledtabular}
			\begin{tabular}{r r r r r r r r r r r}
$B$ (T)	&	0 K	&	300 K	&&	$B$ (T)	&	0 K	&	300 K	&&	$B$ (T)	&	0 K	&	300 K	\\
\hline
0	&	-168.552041	&	-168.552041	&&	8	&	-168.533899	&	-168.535348	&&	160	&	-161.809395	&	-162.242261	\\
0.1	&	-168.552037	&	-168.552039	&&	9	&	-168.529081	&	-168.530915	&&	170	&	-160.998758	&	-161.473738	\\
0.2	&	-168.552030	&	-168.552032	&&	10	&	-168.523698	&	-168.525961	&&	180	&	-160.150516	&	-160.667960	\\
0.3	&	-168.552015	&	-168.552019	&&	20	&	-168.438787	&	-168.447803	&&	190	&	-159.265944	&	-159.826064	\\
0.4	&	-168.551995	&	-168.552000	&&	30	&	-168.297672	&	-168.317802	&&	200	&	-158.346277	&	-158.949161	\\
0.5	&	-168.551969	&	-168.551977	&&	40	&	-168.100937	&	-168.136347	&&	220	&	-156.406369	&	-157.094633	\\
0.6	&	-168.551938	&	-168.551948	&&	50	&	-167.849370	&	-167.903971	&&	240	&	-154.339764	&	-155.112661	\\
0.7	&	-168.551901	&	-168.551913	&&	60	&	-167.543935	&	-167.621338	&&	260	&	-152.154735	&	-153.011013	\\
0.8	&	-168.551859	&	-168.551874	&&	70	&	-167.185740	&	-167.289235	&&	280	&	-149.858881	&	-150.796930	\\
0.9	&	-168.551809	&	-168.551830	&&	80	&	-166.776017	&	-166.908547	&&	300	&	-147.459176	&	-148.477134	\\
1	&	-168.551757	&	-168.551780	&&	90	&	-166.316080	&	-166.480251	&&	320	&	-144.962011	&	-146.057853	\\
2	&	-168.550907	&	-168.550998	&&	100	&	-165.807310	&	-166.005392	&&	340	&	-142.373248	&	-143.544846	\\
3	&	-168.549489	&	-168.549693	&&	110	&	-165.251129	&	-165.485072	&&	360	&	-139.698262	&	-140.943439	\\
4	&	-168.547504	&	-168.547868	&&	120	&	-164.648977	&	-164.920434	&&	380	&	-136.941989	&	-138.258552	\\
5	&	-168.544953	&	-168.545520	&&	130	&	-164.002303	&	-164.312647	&&	400	&	-134.108968	&	-135.494735	\\
6	&	-168.541835	&	-168.542651	&&	140	&	-163.312539	&	-163.662897	&&	500	&	-118.928095	&	-120.628607	\\
			\end{tabular}
		\end{ruledtabular}
	\end{table}
\end{center}

\begin{center}
	\begin{table}[H]
		\caption{\label{tabS2}Energies of $2s$ states (meV) with different magnetic intensity $B$~(Tesla) at temperature 0~K and 300~K for $r_0$ = 4.2086~nm, $\mu$ = 0.2039~$m_e$, $\kappa = 4.5$.}
		\begin{ruledtabular}
			\begin{tabular}{r r r r r r r r r r r}
$B$ (T)	&	0 K	&	300 K	&&	$B$ (T)	&	0 K	&	300 K	&&	$B$ (T)	&	0 K	&	300 K	\\
\hline
0	&	-38.553893	&	-38.553893	&&	8	&	-38.240316	&	-38.199502	&&	160	&	25.940847	&	27.459482	\\
0.1	&	-38.553843	&	-38.553837	&&	9	&	-38.157907	&	-38.106499	&&	170	&	31.831398	&	33.306752	\\
0.2	&	-38.553695	&	-38.553669	&&	10	&	-38.066222	&	-38.003093	&&	180	&	37.819418	&	39.245494	\\
0.3	&	-38.553448	&	-38.553390	&&	20	&	-36.669754	&	-36.436785	&&	190	&	43.897110	&	45.269125	\\
0.4	&	-38.553102	&	-38.552998	&&	30	&	-34.506863	&	-34.043385	&&	200	&	50.057588	&	51.371775	\\
0.5	&	-38.552657	&	-38.552495	&&	40	&	-31.719080	&	-31.011684	&&	220	&	62.603138	&	63.793606	\\
0.6	&	-38.552114	&	-38.551880	&&	50	&	-28.421034	&	-27.487797	&&	240	&	75.414636	&	76.474757	\\
0.7	&	-38.551471	&	-38.551153	&&	60	&	-24.700889	&	-23.574784	&&	260	&	88.458650	&	89.385396	\\
0.8	&	-38.550730	&	-38.550314	&&	70	&	-20.626315	&	-19.344546	&&	280	&	101.707813	&	102.500637	\\
0.9	&	-38.549890	&	-38.549364	&&	80	&	-16.249938	&	-14.848486	&&	300	&	115.139432	&	115.799484	\\
1	&	-38.548951	&	-38.548302	&&	90	&	-11.613374	&	-10.124672	&&	320	&	128.734481	&	129.264053	\\
2	&	-38.534134	&	-38.531539	&&	100	&	-6.750083	&	-5.202292	&&	340	&	142.476846	&	142.878985	\\
3	&	-38.509466	&	-38.503636	&&	110	&	-1.687376	&	-0.104429	&&	360	&	156.352755	&	156.630985	\\
4	&	-38.474989	&	-38.464645	&&	120	&	3.552157	&	5.150194	&&	380	&	170.350347	&	170.508473	\\
5	&	-38.430758	&	-38.414637	&&	130	&	8.949610	&	10.546065	&&	400	&	184.459323	&	184.501297	\\
6	&	-38.376843	&	-38.353703	&&	140	&	14.488990	&	16.070108	&&	500	&	256.394876	&	255.914993	\\
			\end{tabular}
		\end{ruledtabular}
	\end{table}
\end{center}

\begin{center}
	\begin{table}[H]
		\caption{\label{tabS3}Energies of $2p_{-1}$ states (meV) with different magnetic intensity $B$~(Tesla) at temperature 0~K and 300~K for $r_0$ = 4.2086~nm, $\mu$ = 0.2039~$m_e$, $\kappa = 4.5$.}
		\begin{ruledtabular}
			\begin{tabular}{r r r r r r r r r r r}
$B$ (T)	&	0 K	&	300 K	&&	$B$ (T)	&	0 K	&	300 K	&&	$B$ (T)	&	0 K	&	300 K	\\
\hline
0	&	-49.781570	&	-49.781570	&&	8	&	-49.695186	&	-49.807901	&&	160	&	-14.823995	&	-20.622206	\\
0.1	&	-49.782364	&	-49.782378	&&	9	&	-49.663271	&	-49.807384	&&	170	&	-11.440536	&	-17.322395	\\
0.2	&	-49.783110	&	-49.783167	&&	10	&	-49.626753	&	-49.805961	&&	180	&	-7.987731	&	-13.938192	\\
0.3	&	-49.783809	&	-49.783937	&&	20	&	-49.017308	&	-49.709812	&&	190	&	-4.470652	&	-10.476915	\\
0.4	&	-49.784460	&	-49.784689	&&	30	&	-47.998661	&	-49.363442	&&	200	&	-0.893804	&	-6.944978	\\
0.5	&	-49.785063	&	-49.785423	&&	40	&	-46.624801	&	-48.677958	&&	220	&	6.423547	&	0.308947	\\
0.6	&	-49.785619	&	-49.786139	&&	50	&	-44.946648	&	-47.636100	&&	240	&	13.936854	&	7.785739	\\
0.7	&	-49.786127	&	-49.786838	&&	60	&	-43.007928	&	-46.258786	&&	260	&	21.623704	&	15.455403	\\
0.8	&	-49.786587	&	-49.787521	&&	70	&	-40.844855	&	-44.580043	&&	280	&	29.465547	&	23.293827	\\
0.9	&	-49.787000	&	-49.788186	&&	80	&	-38.487089	&	-42.635354	&&	300	&	37.446831	&	31.281380	\\
1	&	-49.787365	&	-49.788835	&&	90	&	-35.958943	&	-40.457323	&&	320	&	45.554375	&	39.401882	\\
2	&	-49.788396	&	-49.794511	&&	100	&	-33.280459	&	-38.074472	&&	340	&	53.776897	&	47.641857	\\
3	&	-49.784671	&	-49.798912	&&	110	&	-30.468298	&	-35.511273	&&	360	&	62.104653	&	55.989959	\\
4	&	-49.776198	&	-49.802285	&&	120	&	-27.536427	&	-32.788581	&&	380	&	70.529166	&	64.436547	\\
5	&	-49.762993	&	-49.804804	&&	130	&	-24.496664	&	-29.924152	&&	400	&	79.043007	&	72.973350	\\
6	&	-49.745074	&	-49.806567	&&	140	&	-21.359086	&	-26.933141	&&	500	&	122.746523	&	116.789306	\\
			\end{tabular}
		\end{ruledtabular}
	\end{table}
\end{center}

\begin{center}
	\begin{table}[H]
		\caption{\label{tabS4}Energies of $2p_{+1}$ states (meV) with different magnetic intensity $B$~(Tesla) at temperature 0~K and 300~K for $r_0$ = 4.2086~nm, $\mu$ = 0.2039~$m_e$, $\kappa = 4.5$.}
		\begin{ruledtabular}
			\begin{tabular}{r r r r r r r r r r r}
$B$ (T)	&	0 K	&	300 K	&&	$B$ (T)	&	0 K	&	300 K	&&	$B$ (T)	&	0 K	&	300 K	\\
\hline
0	&	-49.781570	&	-49.781570	&&	8	&	-49.564342	&	-49.627092	&&	160	&	-12.207129	&	-15.179979	\\
0.1	&	-49.780728	&	-49.780743	&&	9	&	-49.516073	&	-49.592434	&&	170	&	-8.660116	&	-11.759609	\\
0.2	&	-49.779839	&	-49.779895	&&	10	&	-49.463199	&	-49.554019	&&	180	&	-5.043756	&	-8.263017	\\
0.3	&	-49.778902	&	-49.779027	&&	20	&	-48.690200	&	-48.961283	&&	190	&	-1.363123	&	-4.695704	\\
0.4	&	-49.777917	&	-49.778139	&&	30	&	-47.507998	&	-48.000679	&&	200	&	2.377279	&	-1.062559	\\
0.5	&	-49.776885	&	-49.777231	&&	40	&	-45.970585	&	-46.702090	&&	220	&	10.021739	&	6.384172	\\
0.6	&	-49.775806	&	-49.776300	&&	50	&	-44.128877	&	-45.101426	&&	240	&	17.862153	&	14.047174	\\
0.7	&	-49.774678	&	-49.775347	&&	60	&	-42.026603	&	-43.234227	&&	260	&	25.876112	&	21.901854	\\
0.8	&	-49.773503	&	-49.774373	&&	70	&	-39.699976	&	-41.132761	&&	280	&	34.045063	&	29.927753	\\
0.9	&	-49.772280	&	-49.773376	&&	80	&	-37.178656	&	-38.825110	&&	300	&	42.353455	&	38.107629	\\
1	&	-49.771009	&	-49.772357	&&	90	&	-34.486955	&	-36.335246	&&	320	&	50.788108	&	46.426794	\\
2	&	-49.755685	&	-49.760828	&&	100	&	-31.644917	&	-33.683487	&&	340	&	59.337737	&	54.872614	\\
3	&	-49.735604	&	-49.746637	&&	110	&	-28.669202	&	-30.887043	&&	360	&	67.992602	&	63.434133	\\
4	&	-49.710777	&	-49.729480	&&	120	&	-25.573778	&	-27.960544	&&	380	&	76.744223	&	72.101781	\\
5	&	-49.681216	&	-49.709113	&&	130	&	-22.370460	&	-24.916482	&&	400	&	85.585173	&	80.867150	\\
6	&	-49.646942	&	-49.685347	&&	140	&	-19.069329	&	-21.765588	&&	500	&	130.924230	&	125.926405	\\			
			\end{tabular}
		\end{ruledtabular}
	\end{table}
\end{center}

\begin{center}
	\begin{table}[H]
		\caption{\label{tabS5}Energies of $3s$ states (meV) with different magnetic intensity $B$~(Tesla) at temperature 0~K and 300~K for $r_0$ = 4.2086~nm, $\mu$ = 0.2039~$m_e$, $\kappa = 4.5$.}
		\begin{ruledtabular}
			\begin{tabular}{r r r r r r r r r r r}
$B$ (T)	&	0 K	&	300 K	&&	$B$ (T)	&	0 K	&	300 K	&&	$B$ (T)	&	0 K	&	300 K	\\
\hline
0	&	-16.551620	&	-16.551620	&&	8	&	-15.001325	&	-13.427883	&&	160	&	137.144400	&	140.552814	\\
0.1	&	-16.551360	&	-16.551008	&&	9	&	-14.619501	&	-12.803618	&&	170	&	149.056334	&	152.353631	\\
0.2	&	-16.550579	&	-16.549174	&&	10	&	-14.204125	&	-12.148735	&&	180	&	161.055614	&	164.238685	\\
0.3	&	-16.549277	&	-16.546120	&&	20	&	-8.618289	&	-5.574895	&&	190	&	173.134690	&	176.201768	\\
0.4	&	-16.547456	&	-16.541853	&&	30	&	-1.282591	&	2.143361	&&	200	&	185.286995	&	188.237342	\\
0.5	&	-16.545114	&	-16.536377	&&	40	&	7.125278	&	10.705982	&&	220	&	209.788965	&	212.506621	\\
0.6	&	-16.542253	&	-16.529703	&&	50	&	16.266355	&	19.925459	&&	240	&	234.523093	&	237.012433	\\
0.7	&	-16.538873	&	-16.521841	&&	60	&	25.946909	&	29.681931	&&	260	&	259.458929	&	261.726817	\\
0.8	&	-16.534974	&	-16.512804	&&	70	&	36.044740	&	39.855984	&&	280	&	284.571899	&	286.626551	\\
0.9	&	-16.530558	&	-16.502206	&&	80	&	46.477076	&	50.343774	&&	300	&	309.841859	&	311.692133	\\
1	&	-16.525624	&	-16.489263	&&	90	&	57.184816	&	61.074853	&&	320	&	335.252069	&	336.907025	\\
2	&	-16.448058	&	-16.271898	&&	100	&	68.124039	&	72.004863	&&	340	&	360.788456	&	362.257080	\\
3	&	-16.320131	&	-15.948198	&&	110	&	79.261069	&	83.104559	&&	360	&	386.439062	&	387.730100	\\
4	&	-16.143689	&	-15.545495	&&	120	&	90.569426	&	94.353023	&&	380	&	412.193633	&	413.315496	\\
5	&	-15.921016	&	-15.076549	&&	130	&	102.027862	&	105.734178	&&	400	&	438.043305	&	439.004012	\\
6	&	-15.654647	&	-14.570256	&&	140	&	113.619029	&	117.235016	&&	500	&	568.478171	&	568.738960	\\
			\end{tabular}
		\end{ruledtabular}
	\end{table}
\end{center}

\section{\label{sec:S2}Analysis of thermo-induced effect by the perturbation method}

In the center-of-mass frame of reference, the relative motion between the electron and hole in neutral exciton is described by the partial wave function $\psi_{\mathbf{K}} (\mathbf{r})$ corresponding to the pseudomomentum $\mathbf{K}$. This wavefunction is governed by the following one-particle Schr\"odinger equation
\begin{eqnarray}\label{Seq6}
\biggl\{ \frac{\hat{\mathbf{p}}^2}{2 \mu} + V_{RK}(\mathbf{r})+ V_{Zeeman} \left( \mathbf{r} \right) + V_{diamag} \left( \mathbf{r} \right) + V_{mS} (\mathbf{r}) - E \biggr\} \psi_{\mathbf{K}} (\mathbf{r}) = 0.
\end{eqnarray}
Here, we use the Rytova-Keldysh potential 
\begin{equation}\label{Seq9}
V_{RK}\left( \mathbf{r} \right)=-\frac{e^2}{8\varepsilon_0 r_0} \left[H_0\left(\frac{\kappa \,r}{r_0}\right)
    -Y_0\left(\frac{\kappa \,r}{r_0} \right) \right]
\end{equation}
to describe the electron-hole interaction as in many Refs. \cite{SRytova1967, Skeldysh1979, Sharamura1988, Scudazzo2011, Sberkelbach2013, Schernikov2014}. The next two terms are the Zeeman-splitting and diamagnetic potentials, respectively,
\begin{equation}
V_{Zeeman} \left( \mathbf{r} \right) =  \frac{1-\sigma}{1+\sigma} \frac{{eB}}{2\mu}{{\hat l}_z},\quad\qquad
V_{diamag} \left( \mathbf{r} \right) = \frac{{{e^2}{B^2}}}{{8 \mu }}{\mathbf{r}^2} ,
\end{equation}
while the last term is the thermo-induced motional Stark potential
\begin{eqnarray}\label{Seq6c}
V_{mS} (\mathbf{r}) =- \sqrt{\frac{2k_B T}{M}} eB\, r \cos \varphi .
\end{eqnarray}

For illustration, we show the effective potential of equation \eqref{Seq6} in Fig.~\ref{figS1} for two cases, with and without the thermo-induced motional Stark potential, i.e., $T \neq 0$ and $T=0$. This figure clearly shows that at room temperature, the contribution of the added term on the effective potential is noticeable for the 2s and higher Rydberg states, and it would influence the exciton energy spectra.

The effective potential in Fig.~\ref{figS1} shows no tunneling effect, even considering the thermo-induced motional Stark potential. That means the exciton is always in its bound states, unlike the well-known LoSurdo-Stark effect in the two-dimensional electron gas \cite{STanaka1987, SHenriques2020}. Indeed, this fact can be understood from the thermo-induced motional Stark potential \eqref{Seq6c}. This part is linearly proportional to the variable $x$; however, the main part of the effective potential, the diamagnetic potential $V_{diamag} (\mathbf{r})$, is quadratic proportional to the electron-hole distance and consequently dominant at the large separation of the electron-hole pair that keeps the exciton in bound states. Instead of tunneling, we believe the thermo-induced motional Stark potential cause the Stark shift in the energy spectra, as observed in numerical calculations in Subsection \ref{sec:S1C}.

Now we analyze energy spectra of neutral exciton in monolayer TMDC monolayer, solving the Schr{\"o}dinger equation \eqref{Seq6} in the low and high limits of magnetic field intensity by the perturbation theory.

\subsection{\label{sec:S2A}In low magnetic field intensity limit}

The Schr\"odinger equation \eqref{Seq6} without a magnetic field ($B =0$) is much simpler as
\begin{equation}\label{Seqn:zeroth}
\left\{ \frac{\hat{\mathbf{p}}^2}{2 \mu} + V_{RK}(\mathbf{r}) - E^{(0)} \right\} \psi^{(0)} (\mathbf{r}) = 0
\end{equation}
because there are no Zeeman-splitting, diamagnetic, and even thermo-induced motional Stark potentials. Consequently, the energy spectrum and eigenstate of neutral exciton are found as $E_{nm}^{(0)}$ and $\psi_{nm}^{(0)}(\mathbf{r}) =  R_{nm}^{(0)} (r) \dfrac{e^{i m \varphi}}{\sqrt{2\pi}}$, respectively. Here, $m$ is the magnetic quantum number associated with the angular momentum $\hat{l}_z$; and $n = n_r +|m|+1$ is defined as the principal quantum number, with $n_r$ being the node number of the radial wavefunction $R_{nm}^{(0)} (r)$. At the state $\left|n m\right>$, the average separation between electron and hole is $\left< r \right>_{nm} = \left< R_{n m}^{(0)} \right| r \left| R_{nm}^{(0)} \right>$. We note that $E_{nm}^{(0)}$ and $R_{nm}^{(0)} (r)$ are obtained by numerical calculation, particularly by the method described in Sec.~\ref{FKMethod}.

Turning on the magnetic field ($B >0$) such that its typical length is much larger than the average exciton radius, $l_B = \sqrt{\frac{\hbar}{e B}} \gg \left< r \right>_{nm}$, all the Zeeman-splitting, diamagnetic, and thermo-induced motional Stark potentials can be treated as perturbed interactions besides the zeroth-order Hamiltonian in Eq.~\eqref{Seqn:zeroth}. Both the Zeeman-splitting and diamagnetic potentials contribute to the first-order correction via Zeeman-splitting and diamagnetic energies
\begin{eqnarray}
&& \left< \psi_{nm}^{(0)} \right| V_{Zeeman} \left| \psi_{nm}^{(0)} \right> = \frac{1 - \sigma}{1 + \sigma} \frac{\hbar m}{2 \mu} e B, \\
&&\left< \psi_{nm}^{(0)} \right| V_{diamag} \left| \psi_{nm}^{(0)} \right> = \frac{e^2 B^2}{8 \mu} \left< r^2 \right> _{nm}.
\end{eqnarray}

Since the original state $\psi_{nm}^{(0)}(\mathbf{r})$ exhibits the $SO(2)$ symmetry of neutral exciton and the thermo-induced motional Stark potential $V_{mS} (\mathbf{r})$ is proportional to $r \cos \varphi$, its first-order correction must vanish
\begin{equation}
\left< \psi_{nm}^{(0)} \right| V_{mS} \left| \psi_{nm}^{(0)} \right> = 0.
\end{equation}
However, the thermo-induced motional Stark effect solely appears in the second-order correction of energy as
\begin{equation}
\sum_{(n'm')\neq (nm)}\frac{\left|\left< \psi_{nm}^{(0)} \right| V_{mS} \left| \psi_{n'm'}^{(0)} \right> \right|^2}{E_{nm}^{(0)} - E_{n'm'}^{(0)}} = - \alpha_{nm} \frac{k_B T}{M}  e^2 B^2.
\end{equation}
Here, $\alpha_{nm}$ is the zero-field polarizability of the exciton
\begin{eqnarray}
\label{eqn9b}
\alpha_{nm} =  - \sum\limits_{\substack{n' \geq |m'|+1 \\ m' = m\pm 1}} \frac{\left| \int\limits_0^{+\infty} R_{nm}^{(0) \; *}(r)  R_{n'm'}^{(0)} (r) r^2 dr \right|^2}{E_{nm}^{(0)} - E_{n'm'}^{(0)}}.
\end{eqnarray}

Finally, the energy of magnetoexciton at the state $\left|nm\right>$ is obtained as
\begin{eqnarray}\label{Seqn9a}
E_{nm}^{(2)} (B, T)  \approx E_{nm}^{(0)}  + \frac{1 - \sigma}{1 + \sigma} \frac{\hbar  m}{2 \mu} e B + \frac{e^2 B^2}{8 \mu} \left< r^2 \right> _{nm} - \alpha_{nm} \frac{k_B T}{M}  e^2 B^2\,.
\end{eqnarray}
This is the equation (4) in the main text. In other words, the quadratic thermo-induced motional Stark correction in the low magnetic field is proportional to the square of magnetic field strength
\begin{equation}\label{Seqn2}
\Delta E_{nm} (B, T) = E_{nm}^{(2)} (B, T) - E_{nm} (B, 0) \approx \alpha_{nm} \frac{k_B T}{M}  e^2 B^2.
\end{equation}

From energies of the states $1s,\, 2s,\, 3s,\, 2p_{-1}$, and \,$2p_{+1}$ in Tables \ref{tabS1}, \ref{tabS2}, \ref{tabS3}, \ref{tabS4}, and \ref{tabS5}, the quadratic thermo-induced motional Stark corrections in the low magnetic field $\Delta E_{nm} (B, T)$ for these states are numerically determined. Consequently, the zero-field polarizability of exciton ${{\alpha }_{nm}}$ and the thermo-induced Stark correction on diamagnetic coefficient $\Delta \sigma_{nm}=- \frac{ e^2}{M} \alpha_{nm} k_B T$ are revealed by fitting the data of $\Delta E_{nm} (B, T)$ versus ${{B}^{2}}$ in Eq. \eqref{Seqn2}. These results are presented in Table 2 of the main text.

\begin{figure}[H]
\begin{center}
\includegraphics[width=0.49 \columnwidth]{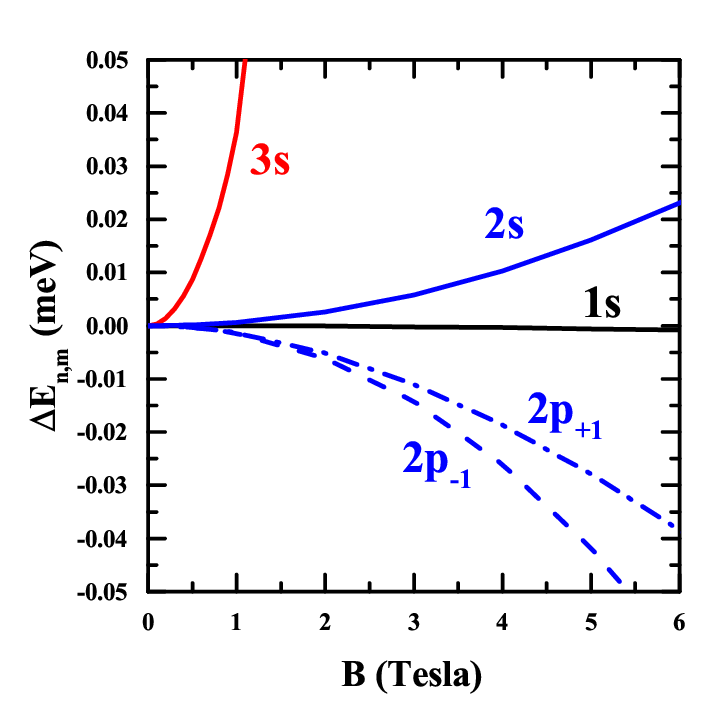}
\caption{The dependence of quadratic thermo-induced motional Stark corrections $\Delta E_{nm} (B, T)$ for various states: $1s$ (black), $2s$ (red), $3s$ (blue), $2p_{-1}$ (blue dashed), and $2p_{+1}$ (blue dot-dashed) on magnetic field strength $B$ for the low magnetic field regime.}
\label{figS2}
\end{center}
\end{figure}

Figure \ref{figS2} indicates that the quadratic thermo-induced motional Stark corrections of $1s,\, 2s,\, 3s,\, 2p_{-1}$, and $ 2p_{+1}$ states depend on ${{B}^{2}}$ for the low magnetic field regime.

\subsection{\label{sec:S2B}In high magnetic field intensity limit}

Suppose the magnetic field is so large that its typical length is much smaller than the average exciton radius, $l_B  \ll \left< r \right>_{nm}^{(0)}$; the diamagnetic potential $V_{diamag} = \frac{e^2 B^2}{8 \mu} \mathbf{r}^2$ (the harmonic oscillator potential of frequency $\omega _B = {e B}/{2\mu}$) becomes dominant. In this case, the energy spectrum tends to the conventional Landau levels. Within the effect of Zeeman-splitting potential $V_{Zeeman}$, these zero-temperature Landau levels are shifted as
\begin{equation}
E_{nm}^{(0)}(B \to \infty, T = 0) = \frac{\hbar}{2 \mu} \left( 2 n - |m| - 1 + \frac{1-\sigma}{1+\sigma} m  \right) e B, 
\end{equation}
and the corresponding eigenstate is of the isotropic harmonic oscillator with the wave function 
\begin{equation}\label{Seqn:high}
\psi_{nm}^{(0)}(\mathbf{r}) = \frac{1}{l_B} R_{nm}^{(0)}\left( \frac{r^2}{l_B^2} \right) \frac{e^{i m \varphi}}{\sqrt{2\pi}},
\end{equation}
where $R_{nm}^{(0)}$ is the dimensionless radial wave function of the dimensionless variable ${r^2}/{l_B^2}$.

Similar to the low magnetic field limit in the finite temperature, the thermo-induced motional Stark potential in the high magnetic field limit contributes to the second-order correction in the perturbation theory as 
\begin{equation}
\sum_{(n',m')\neq (n,m)}\frac{\left|\left< \psi_{nm}^{(0)} \right| V_{mS} \left| \psi_{n'm'}^{(0)} \right> \right|^2}{E_{nm}^{(0)} - E_{n'm'}^{(0)}}.
\end{equation}
It results in the energy spectrum of magnetoexciton behaving as
\begin{eqnarray}
\label{Seqn9c}
E_{nm}^{(2)} (B, T) \approx \frac{\hbar}{2 \mu} \left( 2 n - |m| - 1 + \frac{1-\sigma}{1+\sigma} m  \right) e B - \frac{8 \mu \beta_{nm}}{M} k_B T,
\end{eqnarray}
where $\beta_{nm}$ is a dimensionless constant calculated as
\begin{eqnarray}
\label{eqn9d}
\beta_{nm} && = - \frac{e^2 B^2}{4 \mu} \sum_{(n',m')\neq (n,m)}\frac{\left|\left< \psi_{nm}^{(0)} \right| r \cos\varphi \left| \psi_{n'm'}^{(0)} \right> \right|^2}{E_{nm}^{(0)} - E_{n'm'}^{(0)}} \nonumber\\
&& = - \sum\limits_{\substack{n' \geq |m'|+1 \\ m' = m\pm 1}} \frac{\frac{1}{4}\left| \int\limits_0^{+\infty} R_{nm}^{(0) \; *} (x)  R_{n'm'}^{(0)} (x) \sqrt{x} dx \right|^2}{2(n-n')-(|m|-|m'|)+\frac{1-\sigma}{1+\sigma}(m-m')}.
\end{eqnarray}
Therefore, the quadratic thermo-induced motional Stark correction in the high magnetic field limit solely depends on temperature as
\begin{equation}
\Delta E_{nm} (B, T) = E_{nm}^{(2)} (B, T) - E_{nm} (B, 0) \approx -\frac{8 \mu \beta_{nm}}{M} k_B T.
\end{equation}

Although these calculations are only valid for extremely high magnetic intensity, the magnetic independence of $\Delta E_{nm} (B, T)$ can still be observed in our numerical calculations. Figure \ref{figS3} below illustrates how the quadratic thermo-induced motional Stark correction $\Delta E_{nm} (B, T)$ constantly behaves in the high magnetic field regime.

\begin{figure}[H]
\begin{center}
\includegraphics[width=0.49 \columnwidth]{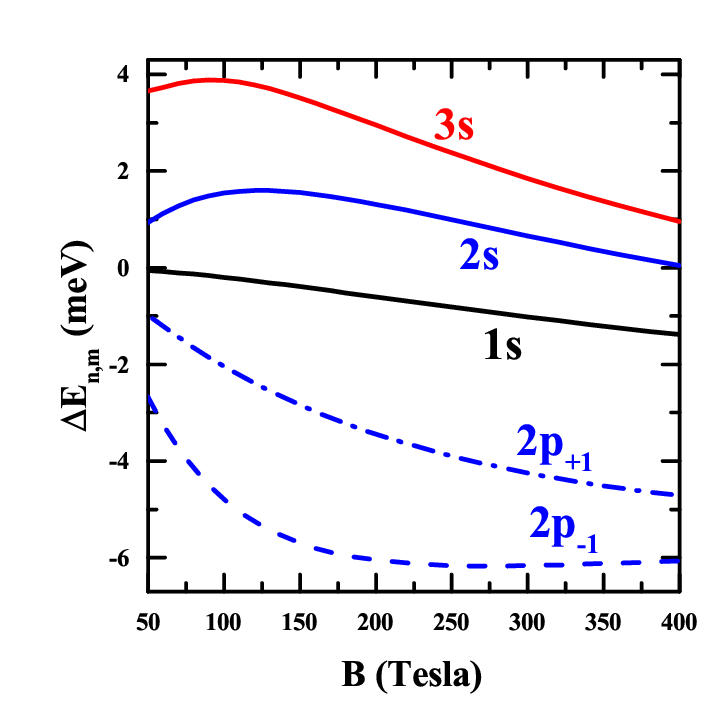}
\caption{The dependence of quadratic thermo-induced motional Stark correction $\Delta E_{nm} (B, T)$ for various states: $1s$ (black), $2s$ (red), $3s$ (blue), $2p_{-1}$ (blue dashed), and $2p_{+1}$ (blue dot-dashed) on magnetic field strength $B$ for the high magnetic field regime.}
\label{figS3}
\end{center}
\end{figure}

\subsection{\label{sec:S2C}Thermo-induced deformation of magnetoexciton wave functions by symmetry-breaking mechanism}

The influence of thermo-induced motional Stark potential on the wave function of magnetoexciton is also important because this potential breaks the $SO(2)$ symmetry of magnetoexciton. Indeed, the angular momentum $\hat{l}_z$ is no longer conserved when we turn on the thermo-induced motional Stark potential. Hence, we can witness the deformation of magnetoexciton's orbital from this breaking symmetry mechanism.

In detail, the relative Hamiltonian of electron-hole pair under the presence of the thermo-induced motional Stark potential reads
\begin{eqnarray}\label{Seqn:H-rel-S2C}
{{\hat H}_{rel}} (T) 
= {{\hat H}_{rel}} (0) + V_{mS} \left( \mathbf{r} \right),
\end{eqnarray}
where the zero-temperature relative Hamiltonian is given as
\begin{eqnarray}\label{Seqn:H-rel-S2C-1}
{{\hat H}_{rel}} (0) &=& \frac{1}{{2\mu }}{{\hat {\mathbf p}}^2} +\frac{1-\sigma}{1+\sigma}\,\frac{{eB}}{2\mu}\,{\hat l}_z +
\frac{{{e^2}{B^2}}}{{8\mu }}{r^2} + V_{RK} \left( r \right) \nonumber\\
&=& - \frac{\hbar^2}{2 \mu} \left[ \frac{1}{r} \frac{\partial}{\partial r} \left( r \frac{\partial}{\partial r} \right) \right] + \frac{\hat{l}_z^2}{2 \mu} +\frac{1-\sigma}{1+\sigma}\,\frac{{eB}}{2\mu}\,{\hat l}_z +
\frac{{{e^2}{B^2}}}{{8\mu }}{r^2} +  V_{RK} \left( r \right).
\end{eqnarray}
One can easily verify that each term of the zero-temperature relative Hamiltonian \eqref{Seqn:H-rel-S2C-1} commutes with the relative angular momentum $\hat{l}_z = -i \hbar \partial / \partial \varphi$. The first, fourth, and fifth terms in Eq. \eqref{Seqn:H-rel-S2C-1} commute with $\hat{l}_z$ because they only depend on the radial variable $r$ and do not depend on the angle variable $\varphi$. On the other hand, the second and third terms are polynomials of $\hat{l}_z$ by themself and trivially commute with $\hat{l}_z$. Then
\begin{equation}
\label{eqn:H-rel-S2C-3}
\left[ \hat{l}_z , {{\hat H}_{rel}} (0) \right] = 0,
\end{equation}    
i.e., the angular momentum $\hat{l}_z$ is conserved at zero temperature $T = 0$. In other words, the magnetoexciton at zero temperature exhibits the $SO(2)$ symmetry. A direct consequence of this $SO(2)$ symmetry is that the wavefunction of the relative motion of magnetoexciton must be in the following form
\begin{equation}
\psi_{nm}^{(0)}(\mathbf{r}) =  R_{nm}^{(0)} (r) \dfrac{e^{i m \varphi}}{\sqrt{2\pi}},
\end{equation} 
where its modulus does not depend on the angle variable $\varphi$:
\begin{equation}
\left|\psi_{nm}^{(0)}(\mathbf{r})\right|^2 =  \left|R_{nm}^{(0)} (r) \right|^2.
\end{equation} 
We say that magnetoexciton at zero temperature has circular orbitals.

The scenario changes when we turn on the temperature $T \neq 0$. Including the thermo-induced motional Stark potential $V_{mS} (\mathbf{r}) =- \sqrt{\frac{2k_B T}{M}} eB r \cos \varphi$ makes the angular momentum $\hat{l}_z$ no longer commutes with the Hamiltonian
\begin{eqnarray}
\left[ \hat{l}_z , {{\hat H}_{rel}} (T) \right] &&= \left[ \hat{l}_z , {{\hat H}_{rel}} (0) \right] + \left[ \hat{l}_z , V_{mS} (\mathbf{r})) \right] \nonumber\\
&&= \left[ - i \hbar \frac{\partial}{\partial \varphi} , - \sqrt{\frac{2k_B T}{M}} eB r \cos \varphi \right] = - i \hbar \sqrt{\frac{2k_B T}{M}} eB r \sin \varphi \neq 0.
\end{eqnarray}
We say that the thermo-induced motional Stark potential breaks the $SO(2)$ symmetry of magnetoexciton. Therefore, it is expected that the thermo-induced motional Stark effect could deform the magnetoexciton's orbitals and change the average distance of the electron-hole pair. We consider the zero-temperature magnetoexciton as zeroth-order Hamiltonian and then treat the thermo-induced motional Stark potential as a perturbation term. The zeroth-order Schr\"odinger equation for magnetoexciton reads
\begin{eqnarray}\label{Seqn:0-general}
\biggl\{ \frac{\hat{\mathbf{p}}^2}{2 \mu} + V_{RK}(\mathbf{r})+ V_{Zeeman} \left( \mathbf{r} \right) + V_{diamag} \left( \mathbf{r} \right) - E^{(0)} \biggr\} \psi_{\mathbf{K}}^{(0)} (\mathbf{r}) = 0.
\end{eqnarray}
The energy spectrum and eigenstates of magnetoexciton are $E_{nm}^{(0)}$ and $\psi_{nm}^{(0)}(\mathbf{r}) =  R_{nm}^{(0)} (r) \dfrac{e^{i m \varphi}}{\sqrt{2\pi}}$, respectively.

\begin{figure}[H]
\begin{center}
\includegraphics[width=0.49 \columnwidth]{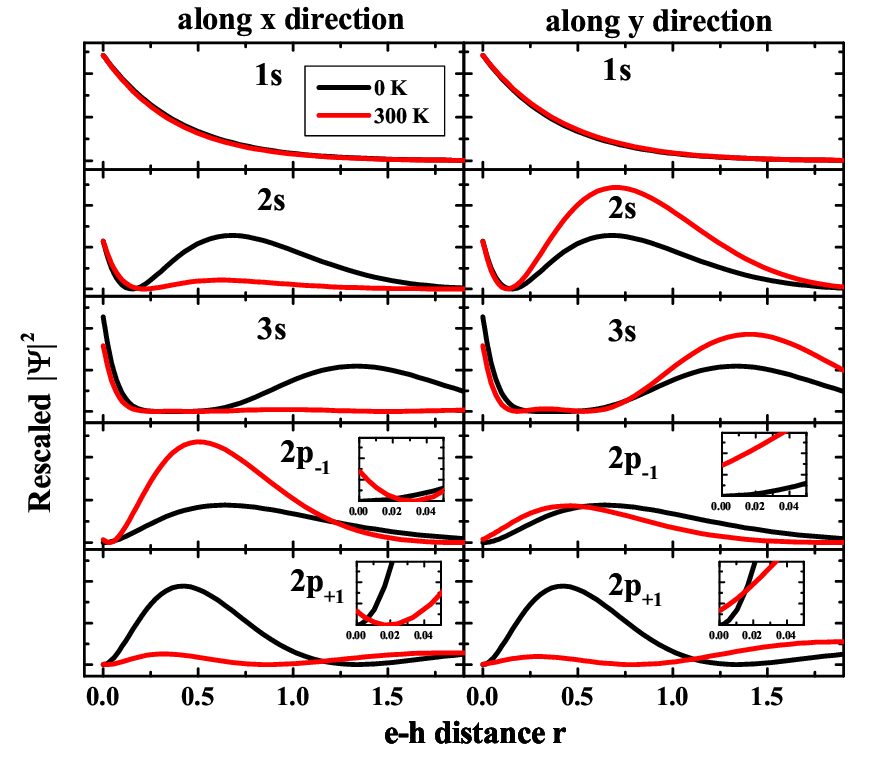}
\caption{Squared modulus of wavefunction $\left| \Psi \right|^2$ as a function of electron-hole separation $r$ along $x$ and $y$ axis for different states.}
\label{figS4}
\end{center}
\end{figure}

Taking into account the first-order correction caused by this thermo-induced motional Stark potential, we obtain the temperature-dependent wave functions of magnetoexciton as
\begin{eqnarray}
\label{eqn9g}
\psi_{nm}^{T}(\mathbf{r})  \approx  R_{nm}^{(0)}(r) \frac{e^{i m \varphi}}{\sqrt{2\pi}} &&  - \sqrt{\frac{2 k_B T}{M}} e B \sum\limits_{n'\geq |m + 1|+1}  f_{nm;n',m+1}R_{n',m+1}^{(0)}(r) \frac{e^{i (m+1) \varphi}}{\sqrt{2\pi}} \nonumber\\
&& - \sqrt{\frac{2 k_B T}{M}} e B \sum\limits_{n'\geq |m-1|+1} f_{nm;n',m-1} R_{n',m-1}^{(0)}(r) \frac{e^{i (m-1) \varphi}}{\sqrt{2\pi}},
\end{eqnarray}
where the coefficients $f_{nm;n'm'}$ are given by the following integrals
\begin{eqnarray}
f_{nm;n'm'} = \frac{\int\limits_{0}^{+\infty} R_{nm}^{(0) \; *}(r) R_{n'm'}^{(0)} (r) r^2 dr}{E^{(0)}_{nm} - E^{(0)}_{n'm'}} .
\end{eqnarray}
As a result, the orbital of magnetoexciton is deformed along the $Ox$ axis because the probability density is no longer angle-independent
\begin{eqnarray}\label{Seqn:deform}
\left| \psi_{nm}^{T} (\mathbf{r}) \right|^2 =  \frac{1}{2 \pi} \left| R_{nm}^{(0)} (r) \right|^2  && - \frac{1}{2 \pi} \sqrt{\frac{2 k_B T}{M}} eB \sum _{n'} \left( f_{nm;n',m+1} R_{nm}^{(0) \; *}(r) R_{n',m+1}^{(0)} (r) e^{i \varphi} \right. \nonumber\\
&& \left. +  f_{nm;n',m-1} R_{nm}^{(0) \; *}(r) R_{n',m-1}^{(0)} (r) e^{-i \varphi} + c.c\right) \nonumber\\
&& + \frac{1}{2 \pi} \frac{2 k_B T}{M} e^2 B^2 \sum\limits_{n',n''} \left[ f^*_{nm;n'',m+1} f_{nm;n',m+1} R_{n'',m+1}^{(0) \; *}(r) R_{n',m+1}^{(0)} (r) \right. \nonumber\\
&& + f^*_{nm;n'',m-1} f_{nm;n',m-1} R_{n'',m-1}^{(0) \; *}(r) R_{n',m-1}^{(0)} (r) \nonumber\\
&& \left. + \left( f^*_{nm;n'',m-1} f_{nm;n',m+1} R_{n'',m+1}^{(0) \; *}(r) R_{n',m-1}^{(0)} (r) e^{-2 i \varphi} + c.c  \right) \right].
\end{eqnarray}
The symmetry breaking is clear in the squared modulus of the wave functions $|{\psi}|^2$ of $2s$, $3s$, $2p_{-1}$ and $2p_{+1}$ states at temperatures of 0K and 300K, as shown in Fig.~\ref{figS4}. The effect is also seen in the orbital of the $1s$ state but is very slight.

\begin{figure}[H]
\begin{center}
\includegraphics[width=0.4 \columnwidth]{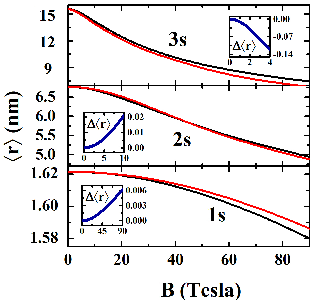}
\caption{Magnetic field dependence of average radius $\left<r \right>$ of magnetoexciton with different states. Small panels show the change of average radius $\Delta \left<r \right>$ by temperature.}
\label{figS5}
\end{center}
\end{figure}

A direct consequence of the orbital's deformation is the change of the average separation between electron and hole in magnetoexcitons
\begin{eqnarray}
\left< r \right>_{nm}^{T} && =  \int_{0}^{+\infty} \int_0^{2 \pi} \left| \psi_{nm}^{(T)} (\mathbf{r}) \right|^2 r^2 dr d\varphi \nonumber\\
&& = \int_0^{+\infty} \left| R_{nm}^{(0)} (r) \right|^2 r^2 dr + \frac{2 k_B T}{M} e^2 B^2 \sum\limits_{\substack{n',n''\\m'=m\pm 1}} f^*_{nm;n''m'} f_{nm;n'm'} \int_0^{+\infty}  R_{n''m'}^{(0) \; *}(r) R_{n'm'}^{(0)} (r) r^2 dr \nonumber\\
&& = \left< r \right>_{nm} + \gamma _{nm} e^2 B^2 \frac{k_B T}{M},
\end{eqnarray}
with
\begin{eqnarray}
\gamma _{nm} = \sum\limits_{\substack{n',n'' \geq |m'|+1 \\m'=m\pm 1}} f^*_{nm;n''m'} f_{nm;n'm'} \int_0^{+\infty}  R_{n''m'}^{(0) \; *}(r) R_{n'm'}^{(0)} (r) r^2 dr.
\end{eqnarray}
Figure \ref{figS5} shows how the average electron-hole separation $<r>$ depends on the magnetic field at zero and room temperatures. As seen in the small panels, the change $\Delta <r>$ is proportional to $B^2$ for the low magnetic field limit, as predicted by the perturbation theory.

\subsection{\label{sec:S2D}Squared modulus of magnetoexciton wave functions at zero separation of electron-hole pair for $np$ and $nd$ states}

According to Elliot formula (see equation (9) in the main text), the signal of exciton peaks on the absorption spectra mainly depends on the oscillation strength, which is proportional to the squared modulus of the wave function $\left| \psi _{nm} (\mathbf{r} = 0) \right|^2$ at zero distance of the electron-hole pair. 

At zero temperature $T = 0$, the conservation of the angular momentum $\hat{l}_z$, i. e., the $SO(2)$ symmetry of Schr\"odinger equation \eqref{Seqn:0-general}, suggests the form of the radial wave function as follows
\begin{equation}
\label{eqn:0wave}
R_{nm}^{(0)} \propto r^{|m|} e^{-f(r)} u(r). 
\end{equation}
Therefore, except for the $s$-states, the wave function at zero distance $r = 0$ for all other states must vanish
\begin{equation}
\left| \psi_{np}^{(0)}(\mathbf{r} = 0)\right|^2 = \left| \psi_{nd}^{(0)}(\mathbf{r} = 0)\right|^2 = \ldots = 0.
\end{equation}

\begin{figure}[H]
\begin{center}
\includegraphics[width=0.95 \columnwidth]{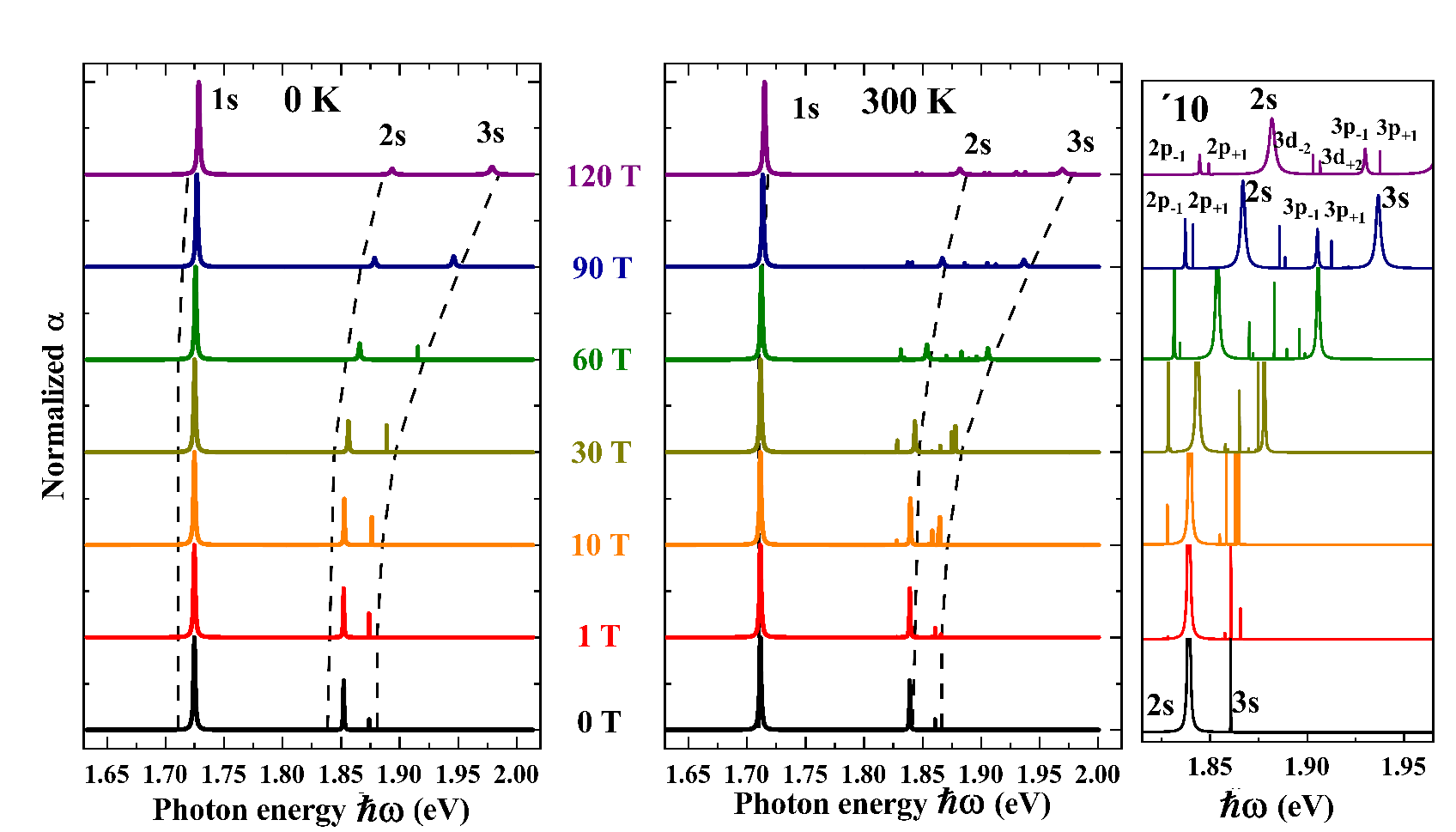}
\caption{Normalized absorption spectra of neutral magnetoexciton at zero temperature (left) and room temperature (center) for various strength of magnetic field. The right panel is offset ($\times 10$ times) near $2s$ and $3s$ peaks of the center panel.}
\label{figS6}
\end{center}
\end{figure}

When we turn on the temperature  ($T \neq 0$), the orbital of magnetoexciton is deformed, as in Eq. \eqref{Seqn:deform} based on the perturbation theory calculations. Hence, for $p$-states, the squared modulus becomes non-vanished
\begin{eqnarray}
\left| \psi_{np}^{T}(\mathbf{r} = 0)\right|^2 = \frac{2 k_B T}{M} e^2 B^2 \sum _{n',n''} f^{*}_{np; n''s} f_{np; n's} \psi_{n''s}^{(0) \; *}(\mathbf{r} = 0) \psi_{n's}^{(0)}(\mathbf{r} = 0) \propto \frac{k_B T}{M}.
\end{eqnarray}
Nonzero wavefunction at zero e-h separation is witnessed for $2p_{\pm 1}$ states as shown in small panels in Fig.~\ref{figS4}. This circumstance leads to the emergence of the $p$-state peaks on the absorption spectra of neutral magnetoexciton, as shown in Fig. 4 of the main text and more details in Fig.~\ref{figS6}.

Our numerical results also highlight the emergence of the $d$-state peaks on the neutral magnetoexciton's absorption spectra. To explain this emergence, we need a higher-order correction beyond the quadratic thermo-induced motional Stark effect. Skipping the detailed calculations, we obtain the squared modulus of magnetoexciton wave function at zero separation for the $d$-state, which is proportional to the temperature square as
\begin{eqnarray}
\left| \psi_{nd}^{T}(\mathbf{r} = 0)\right|^2 \propto \left( \frac{k_B T}{M} \right)^2.
\end{eqnarray}
Thus, the intensity of $d$-states peaks is much smaller than that of the $p$-states.

\end{document}